\documentclass[multphys,vecphys,fleqn,amsmath,amssymb]{svmult}

\usepackage{makeidx}         
\usepackage{graphicx}        
\usepackage{multicol}        
\usepackage[bottom]{footmisc}
\usepackage{bm}

\def\ket#1{\left\vert #1 \right\rangle}

\def\v#1{{\bf #1}}

\def\refeqn#1{Eq.\ (\ref{Equation::#1})}

\def\refeqs#1#2{Eqs.\ (\ref{Equation::#1}) and (\ref{Equation::#2})}

\makeindex             

\begin{document}

\title*{Electron spin as a spectrometer of nuclear spin noise and
  other fluctuations}
\titlerunning{Electron spin as a spectrometer of nuclear spin noise}
\author{Rogerio de Sousa}
\institute{Department of Physics, University of California, Berkeley,
  California 94720, USA\\
\texttt{rdesousa@berkeley.edu}}

\maketitle

\section{Introduction}
\label{intro}

Although the study of electron spin dynamics using pulse electron spin
resonance is an established research field \cite{schweiger01}, many
theoretical questions regarding the microscopic mechanisms for
reversible and irreversible decay of spin coherence remain open.
Recently, the quest towards scalable quantum computation using electron
spins \cite{hu04} gave new impetus to pulse spin resonance, and
sparked major experimental progress towards control and detection of
individual electron spins in the solid state environment
\cite{jelezko04,elzerman04,xiao04,rugar04,petta05}.  

The microscopic understanding of the mechanisms leading to electron
spin energy and phase relaxation, and the question of how to control
these processes is central to the research effort in spin-based
quantum computation.  The goal of theory is to achieve microscopic
understanding so that spin coherence can be controlled either from a
materials perspective (i.e., choosing the best nanostructure for spin
manipulation and dynamics) or from the design of efficient pulse
sequences that reach substantial coherence enhancement without a
high overhead in number of pulses and energy deposition. (The latter is
particularly important in the context of low temperature experiments
where undesired heating from the microwave excitations must be
avoided).

One has to be careful in order to distinguish the time scales
characterizing electron spin coherence. It is customary to introduce
three time scales, $T_1$, $T_2^{*}$, and $T_2$. For localized electron
spins, these time scales usually differ by many orders of magnitude,
because each is dominated by a different physical process.  $T_1$ is
the $1/e$ decay time for the spin magnetization along the external
magnetic field direction.  As an example, $T_1$ of a phosphorus donor
impurity in silicon is of the order of a thousand seconds at low
temperatures and moderate magnetic fields ($T=4$~K and $B=0.3$~T)
\cite{feher59b}.  These long $T_1$'s are explained by noting that the
spin-orbit interaction produces a small admixture of spin up/down
states; the electron-phonon interaction couples these admixtures
leading to $\frac{1}{T_1}\propto B^5$ at low temperatures
\cite{haseg_roth60,desousa03c}.  $T_1$ is generally long because a
spin-flip in a magnetic field requires energy exchange with the
lattice via phonon emission.  The time scales $T_{2}^{*}$ and $T_2$
are instead related to phase relaxation, and hence do not require
transmission of energy to the lattice.  Here $T_2^{*}$ is the 1/e
decay time of the precessing magnetization in a free induction decay
(FID) experiment ($\pi/2-t-\rm{measure}$, where $\pi/2$ denotes a spin
rotation around the $x$ axis). Hence $T_2^{*}$ is the decay time of
the total in-plane magnetization of an ensemble of spins separated in
space or time (e.g., a group of impurities separated in space, or the
time-averaged magnetization of a single spin, as discussed in section
\ref{singlevsmany} below).  For a phosphorus impurity in natural
silicon, $T_2^*\approx~20$~ns due to the distribution of frozen
hyperfine fields, that are time-independent within the measurement
window of the experiment.  
The $T_2^{*}$ decay is
reversible, because the ensemble in-plane magnetization is almost completely
recovered by applying a spin-echo pulse sequence.  In this review we
define $T_2$ as the 1/e decay time of a Hahn echo
($\pi/2-\tau-\pi-\tau-\rm{echo}$).  The irreversible decoherence
time $T_2$ is caused by uncontrolled time dependent fluctuations
within each time interval $\tau$. For a phosphorus impurity in natural
silicon, we have $T_2\approx 0.3$~ms \cite{chiba72}, four orders of
magnitude longer than $T_{2}^{*}$.

The discussion above clearly indicates that the resulting coherence
times are critically dependent on the particular pulse sequence chosen
to probe spin dynamics.  In section \ref{section_noise} we show that
spin coherence can be directly related to the spectrum of electron
spin phase fluctuations and a filter function appropriate for the
particular pulse sequence.  The phase of a precessing electron spin is
a sensitive probe of magnetic fluctuations. This gives us the
opportunity to turn the problem around and view pulse electron spin
resonance as a powerful tool enabling the study of low frequency
magnetic fluctuations arising from complex many-body spin dynamics in the
environment surrounding the electron spin.

A particularly strong source of magnetic noise arises due to the
presence of nuclear spins in the sample.  It is in fact no surprise
that the dominant mechanism for nuclear spin echo \cite{herzog56} and
electron spin echo decay \cite{gordon58,chiba72} has long been related
to the presence of non-resonant nuclear spin species fluctuating
nearby the resonant spin.  Nevertheless, the theoretical understanding
of these experiments was traditionally centered at phenomenological
approaches \cite{klauder62,mims72}, whereby the electron phase
is described as a Markovian stochastic process with free parameters
that can be fitted to experiment (this type of process has been
traditionally denoted spectral diffusion, since the spin resonance
frequency fluctuates along the resonance spectrum in a similar way
that a Brownian particle diffuses in real space).

\begin{figure}[t]
\centerline{\includegraphics*[width=.7\textwidth]{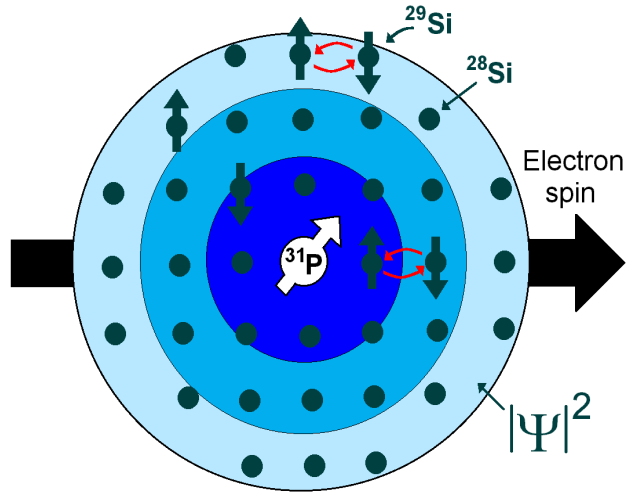}}
\caption[]{The electron spin of a donor impurity in silicon is
  sensitive to the magnetic noise produced by nuclear spins within its
  wave function. When two $^{29}$Si isotopes are close to each other,
  their nuclear spin states may flip-flop due to their mutual dipolar
  interaction. These flip flop events produce time dependent
  fluctuations in the electron's hyperfine field, leading to phase
  relaxation and spin echo decay.}
\label{fig_flipflop}
\end{figure}

Recently, we embarked on an effort aimed at understanding the
mechanism of electron spin coherence due to nuclear spins from a fully
microscopic point of view. In Reference \cite{desousa03b} we developed
a semiclassical model for electron spin echo decay based on the
assumption that the relevant nuclear spin dynamics results from pair
``flip-flops'', where the spin of two nuclei located close to each
other is exchanged due to their mutual dipolar interaction
(Fig.~\ref{fig_flipflop}).  The flip-flop processes lead to
fluctuations in the nuclear spin hyperfine field seen by the localized
electron (e.g., a donor impurity or a quantum dot in a semiconductor).
The semiclassical theory is based on the assumption that each
flip-flop can be described by a random telegraph noise process (a
phenomenological assumption), but with relaxation parameters that can
be derived theoretically from a microscopic theory based on the
nuclear spin dipolar evolution.  Therefore this theory describes the
irreversible decay of the effective hyperfine field produced by a pair
of nuclear spins on the electron spin.  Comparison with experiment
\cite{chiba72,tyryshkin03,tyryshkin06,abe04,ferreti05} suggested
reasonable order of magnitude agreement for the 1/e echo decay time
(within a factor of 3) but poor qualitative agreement for the time
dependence of the echo envelope.  The next step was to develop a full
quantum theory for the nuclear spin dynamics affecting the electron
spin.  In reference \cite{witzel05} a cluster expansion method was
developed to calculate echo decay due to the closed-system dynamics of
a group of dipolar coupled nuclear spins, without any stochastic
assumption about the nuclear spin dynamics.  At lowest order in this
cluster expansion the qualitative and quantitative agreement with
experimental data was quite good.

In section \ref{section_nuclear} we develop a fully microscopic theory
for the nuclear spin noise spectrum arising due to pair flip-flops
induced by the inter-nuclear dipolar interaction. This allows us to
give an elegant and simple derivation of the lowest order cluster
expansion results of Ref.  \cite{witzel05} and to interpret these
results from the point of view of non-equilibrium statistical
mechanics. The full noise spectrum is expressed as a sum of
delta-function contributions corresponding to isolated pair flip-flop
transitions.  We then show that irreversibility can be incorporated
into the pair flip-flop processes by adding broadening to these sharp
transitions, in a mean-field like approach.  Using the method of
moments we are able to calculate these broadenings exactly (at
infinite temperature).  We show explicit numerical results for the
noise spectrum affecting a donor impurity in silicon and compare the
improved theory with echo decay experiments in natural
\cite{tyryshkin06} and nuclear-spin-enriched samples \cite{abe04}.

\section{Noise, relaxation, and decoherence}
\label{section_noise}

When the coupling between the spin qubit and the environment is weak,
we may write a linearized effective Hamiltonian of the form
\begin{equation}
{\cal H}(t)=\frac{1}{2}\gamma B \sigma_z + 
\frac{1}{2}\sum_{q=x,y,z}\hat{\eta}_{q}(t)\sigma_q,
\label{ht}
\end{equation}
where $\bm{B}\parallel \hat{z}$ is a static (time independent)
magnetic field, $\gamma$ is a gyromagnetic ratio in units of $(s
G)^{-1}$ (we set $\hbar=1$ so that energy has units of frequency),
$\sigma_q$ for $q=x,y,z$ are the Pauli matrices describing qubit
observables and $\hat{\eta}_q(t)$ represents the environmental 
(bath) degrees of freedom.

\subsection{The Bloch-Wangsness-Redfield master equation}

In order to describe the long-time dynamics we may take the limit
$t\rightarrow \infty$. Such an approximation is appropriate provided
$t\gg \tau_c$, where $\tau_c$ is a typical correlation time for bath
fluctuations (later we will define $\tau_c$ properly and relax the
long time approximation). In this case spin dynamics can be described
by the Bloch-Wangsness-Redfield theory. The average values of the
Pauli operator satisfy a Master equation (for a derivation see
section 5.11 of Ref.~\cite{slichter96})
\begin{equation}    
\frac{d}{dt}\langle \bm{\sigma}\rangle = \gamma\bm{B}\times \langle
\bm{\sigma}\rangle-\frac{1}{T_1}\langle \sigma_z\rangle \bm{\hat{z}}
-\frac{1}{T_{2x}}\langle\sigma_x\rangle \bm{\hat{x}}
-\frac{1}{T_{2y}}\langle\sigma_y\rangle \bm{\hat{y}}, 
\label{meqn}
\end{equation}
with 
\begin{eqnarray}
\frac{1}{T_1}&=&\frac{\pi}{2}\sum_{q=x,y}
\left[\tilde{S}_{q}(+\gamma B)+\tilde{S}_{q}(-\gamma B)\right],\label{T1}\\
\frac{1}{T_{2x}}&=&
\frac{\pi}{2}\left[\tilde{S}_{y}(+\gamma B)+\tilde{S}_{y}(-\gamma B)\right]
+\pi \tilde{S}_{z}(0),\label{T2x}\\
\frac{1}{T_{2y}}&=&\frac{\pi}{2}
\left[\tilde{S}_{x}(+\gamma B)+\tilde{S}_{x}(-\gamma B)\right]
+\pi \tilde{S}_{z}(0).\label{T2y}
\end{eqnarray}
Here the noise spectrum is defined as
\begin{equation}
\tilde{S}_q(\omega)= \frac{1}{2\pi} \int_{-\infty}^{\infty}
\textrm{e}^{i\omega t} \langle \hat{\eta}_q(t)
\hat{\eta}_{q}(0)\rangle dt.
\label{noise}
\end{equation}
In Eqs.~(\ref{T1}),~(\ref{T2x}),~(\ref{T2y}) we assume
$\langle\hat{\eta}_q(t)\hat{\eta}_{q'}(0)\rangle=0$ for $q\neq q'$.
These equations are the generalization of Fermi's golden rule for
coherent evolution.  From Eq.~(\ref{meqn}) we may show that the
coherence amplitude $|\langle \sigma_+\rangle|=|\langle\sigma_x
+i\sigma_y\rangle|/2$ in a FID experiment decays exponentially with a
rate given by
\begin{equation}
\frac{1}{T^{*}_{2}}= \frac{1}{2}\left(\frac{1}{T_{2x}}
+\frac{1}{T_{2y}}\right)=\frac{1}{2 T_1}+\pi \tilde{S}_{z}(0).\label{T2}
\end{equation}
In contrast, $T_1$ is the time scale
for $\langle \sigma_z\rangle$ to approach equilibrium, i.e., $T_1$ is
the energy relaxation time.  According to Eq.~(\ref{T2}) we have
$T^{*}_{2}\leq 2 T_1$.  Note that $T_1$ depends on the noise spectrum only
at frequencies $+\gamma B$ and $-\gamma B$, a statement of energy
conservation.  Positive frequency noise can be interpreted as
processes where the qubit decays from $\uparrow$ to $\downarrow$ and
the environment absorbs an energy quantum $\gamma B$, while negative
frequency noise refers to qubit excitation (from $\downarrow$ to
$\uparrow$) when the environment emits a quantum $\gamma B$.  The
correlation time $\tau_c$ can be loosely defined as the inverse
cut-off for $\tilde{S}_q(\omega)$, i.e., for $\omega\gg 1/\tau_c$ we
may approximate $\tilde{S}_q(\omega)\approx 0$.

The Master equation [Eq.~(\ref{meqn})] leads to a simple exponential
time dependence for all qubit observables. Actually this is not true
in many cases of interest, including the case of a phosphorus impurity
in silicon where this approximation fails completely (for Si:P the
observed free induction decay is approximately
$\exp{[-(t/T_{2}^{*})^2]}$, while the echo can be fitted to
$\exp{[-(2\tau/T_2)^{2.3}]}$).  The problem lies in the fact that the
$t\rightarrow \infty$ assumption averages out finite frequency
fluctuations; note that $T^{*}_{2}$ differs from $T_1$ only via static
noise [$\tilde{S}_{z}(0)$ in Eq.~(\ref{T2})].  A large number of pulse
spin resonance experiments are sensitive to finite frequencies only
[the most notable example is the spin echo, which is able to remove
$\tilde{S}_z(0)$ completely].  Therefore one must develop a theory for
coherent evolution that includes finite frequency fluctuations.
Analytical results can be derived if we have $\hat{\eta}_q=0$ for
$q=x,y$ (the pure dephasing limit) and $\hat{\eta}_z$ is distributed
according to Gaussian statistics. For many realistic problems the pure
dephasing limit turns out to be a good approximation to describe phase
relaxation.  A common situation is that the neglected components
$\tilde{S}_{x,y}(\omega)$ are much smaller than
$\tilde{S}_{z}(\omega)$ for positive frequencies much smaller than the
qubit energy splitting.  For example, in the case of a localized
electron spin in a semiconductor, $\tilde{S}_{x,y}(\omega)\propto
\omega^5$ due to the combination of the spin-orbit and electron phonon
interactions \cite{desousa03c}. As a result,
$\tilde{S}_{x,y}(\omega)\ll \tilde{S}_{z}(\omega)$ as
$\omega\rightarrow 0$. The Gaussian approximation is described below.

\subsection{Finite frequency phase fluctuations and coherence decay in the 
  Gaussian approximation}

In many cases of interest, the environmental variable $\hat{\eta_z}$
is a sum over several dynamical degrees of freedom, and measurement
outcomes for the operator $\hat{\eta}_z$ may assume a continuum of
values between $-\infty$ and $+\infty$.\footnote{An important
  exception is the observation of individual random telegraph noise
  fluctuators in nanostructures (in this case $\hat{\eta}=\pm \eta'$
  assumes only two discrete values). This results in important
  non-Gaussian features in qubit evolution.}  In those situations we
often can resort to the Central limit theorem which states that the
statistics for outcomes $\eta'$ follows a Gaussian distribution,
\begin{equation}
P[\hat{\eta}_z(t)=\eta']=\frac{1}{\sqrt{2\pi\Delta^2}}
\exp{\left(-\frac{\eta'^2}{2\Delta^{2}}\right)},
\label{pt}
\end{equation}
with a stationary (time independent) variance given by
$\Delta^2=\langle \hat{\eta}_{z}^{2}(t)\rangle$.  Here $\langle A
\rangle=\rm{Tr}\{\hat{\rho}_{\rm B}A\}$ is a thermal average taken
over all bath degrees of freedom
($\hat{\rho}_B\propto\textrm{e}^{-{\cal H}_B/k_B T}$ is the canonical
density matrix for the bath). We also assume
$\langle\hat{\eta}_z(t)\rangle=0$, since any constant drift in
the noise can be incorporated in the effective $B$ field.  

Our problem is greatly simplified if we relate the operator
$\hat{\eta}_z$ to a Gaussian stochastic process $\eta'(t)$ in the
following way.  For each time $t$, $\eta'(t)$ corresponds to a
classical random variable, that can be interpreted as the
\emph{outcome of measurements performed by the qubit on the
  environment}. This allows us take averages over the bath states
using Eq.~(\ref{pt}).  Note that the statement ``Gaussian'' noise
refers specifically to the distribution of noise amplitudes, that is
not necessarily related to the spectrum of fluctuations (see below).

Our simplified effective Hamiltonian leads to the following evolution
operator [recall that we set $\hat{\eta}_{x,y}\equiv 0$ in 
Eq.~(\ref{ht})]
 \begin{equation}
{\cal U}_\eta(t,0)=\textrm{e}^{-i\int_{0}^{t}dt' {\cal
    H}(t')}
=\textrm{e}^{-\frac{i}{2}\sigma_z\left[Bt+\int_{0}^{t}\eta'(t')dt'
\right]}
=\textrm{e}^{-\frac{i}{2}\sigma_z\left[Bt+X_\eta(t)\right]},
\end{equation}
where we define 
\begin{equation}
X_{\eta}(t)=\int_{0}^{t} \eta'(t')dt'.
\end{equation}
Here the subscript $\eta$ emphasizes that this operator is a
functional of the trajectory $\eta'(t')$. The effect of the
distribution of trajectories $\eta'$ can be described by assuming the
qubit evolves according to the density
matrix\footnote{This assumption is equivalent to the Kraus
  representation in the theory of open quantum systems.}
\begin{equation}
\rho(t)=\sum_{\eta}p_\eta {\cal U}_\eta (t,0)\rho_0
{\cal U}^{\dag}_{\eta}(t,0),
\label{rhotk}
\end{equation}
where $p_\eta$ denotes the appropriate weight probability for each
environmental trajectory and $\rho_0$ is the $t=0$ density matrix for
the qubit. 
The coherence envelope at
time $t$ averaged over all possible noise trajectories is then
\begin{eqnarray}
\langle\langle \sigma_+ (t)\rangle\rangle&=&
\rm{Tr}\left\{\sigma_+\rho(t)\right\}\nonumber\\
&=&\sum_\eta p_{\eta}\rm{Tr}\left\{{\cal U}^{\dag}_{\eta}(t,0)
\sigma_+{\cal U}_{\eta}(t,0)\rho_0\right\}\nonumber\\
&=&\textrm{e}^{iBt}\sum_{\eta}p_\eta\textrm{e}^{iX_\eta(t)}
\rm{Tr}\left\{\sigma_+\rho_0\right\}\nonumber\\
&=&
\langle\textrm{e}^{iX_\eta(t)}\rangle 
\textrm{e}^{iBt}
\rm{Tr}\left\{\sigma_+\rho_0\right\},
\label{expx0}
\end{eqnarray}
where we used the identity
$\textrm{e}^{i\alpha\sigma_z}\sigma_+
\textrm{e}^{-i\alpha\sigma_z}=\textrm{e}^{2i\alpha}\sigma_+$.
Here the double average $\langle\langle\cdot\rangle\rangle$ denotes a
quantum mechanical average over the qubit basis plus an ensemble
average over the noise trajectories $\eta'(t)$. We can evaluate the
coherence amplitude explicitly by noting that 
the random variable
$X(t)=\int_{0}^{t}\eta'(t')dt'$ is also described by a Gaussian
distribution, but with a time dependent variance given by
$\sigma_t=\langle X^2(t)\rangle$. Therefore we have
\begin{equation}
\langle \textrm{e}^{iX_\eta(t)}\rangle=
\int_{-\infty}^{\infty} \frac{1}{\sqrt{2\pi}\sigma_t}
\textrm{e}^{-\frac{X^2}{2\sigma_t^2}}\textrm{e}^{iX}
=\textrm{e}^{-\frac{1}{2}\sigma_t^2},
\label{expx}
\end{equation}
with $\sigma_t$ given by
\begin{eqnarray} 
\sigma_t^2&=&\int_0^t dt_1 \int_0^t dt_2 \langle
\eta'(t_1)\eta'(t_2)\rangle\nonumber\\
&=&2\int_0^t dt_1 \int_0^{t_1}dt_2 \langle
\eta'(t_1)\eta'(t_2)\rangle\nonumber\\
&=& 2\int_0^t dt' \int_{t'/2}^{t-t'/2} dT \langle \eta'(T+t'/2)
\eta'(T-t'/2)\rangle\nonumber\\
&=&2\int_0^{t} (t-t') S(t') dt'.
\label{sigtcalc}
\end{eqnarray}
Here we introduced the time dependent correlation function
$S(t')=\langle \eta'(T+t'/2)\eta'(T-t'/2)\rangle =\langle
\eta(t')\eta(0)\rangle$, that is independent of a time translation $T$
by virtue of the stationarity assumption. Based on the discussion
above, it is natural to include quantum noise effects by making the
identification 
\begin{equation}
S(t)=\langle \eta'(t)\eta'(0)\rangle\rightarrow
\langle\hat{\eta}_{z}(t)\hat{\eta}_z(0)\rangle.
\end{equation}

It is straightforward to generalize Eqs.~(\ref{expx}) and
(\ref{sigtcalc}) for echo decay. Instead of free evolution (denoted
free induction decay in magnetic resonance), consider the Hahn echo
given by the sequence $\pi/2-\tau-\pi-\tau-\rm{echo}$. 
Here the notation ``$\pi/2$'' denotes a  perfect, instantaneous
90$^{\circ}$ spin rotation around the x-axis (described by the operator
$\textrm{e}^{-\frac{i}{4}\pi\sigma_x}$). The notation ``$-\tau-$''
means the spin is allowed to evolve freely for a time interval
$\tau$. ``$\pi$'' denotes a 180$^{\circ}$ spin rotation around the
x-axis, also referred as a ``$\pi$-pulse'' (this is described by the operator
$\textrm{e}^{-\frac{i}{2}\pi\sigma_x}=-i\sigma_x$). 
The initial
$\pi/2$ pulse prepares the qubit in the state
$\rho_0=|y+\rangle\langle y+|$, after which it is allowed to evolve
freely for time $\tau$, when the $\pi$-pulse 
is applied. After this pulse the qubit is allowed to evolve 
for a time interval $\tau$ again, after which 
the coherence echo is recorded. Hence the evolution operator is
given by
\begin{equation}
{\cal U}_{\rm{Hahn}}(2\tau)={\cal U}(2\tau,\tau)(-i\sigma_x){\cal U}(\tau,0).
\end{equation}
The same procedure leading to Eq.~(\ref{sigtcalc}) is now repeated in
order to calculate the magnitude of the Hahn echo envelope at $t=2\tau$. 
The quantum average is given by 
\begin{eqnarray}
\langle \sigma_+(2\tau)\rangle &=&
\rm{Tr}\left\{ {\cal U}^{\dag}(\tau,0)(i\sigma_x){\cal U}^{\dag}(2\tau,\tau)
\sigma_+ {\cal U}(2\tau,\tau) (-i\sigma_x) {\cal U}(\tau, 0)
\rho_0\right\}\nonumber\\
&=&\rm{Tr}\left\{
{\cal U}^{\dag}(\tau,0)\sigma_x
\textrm{e}^{i\int_{\tau}^{2\tau}dt' [B+\eta'(t')]}
\sigma_+
\sigma_x {\cal U}(\tau, 0)
\rho_0\right\}\nonumber\\
&=&\rm{Tr}\left\{
\textrm{e}^{i\int_{\tau}^{2\tau}dt' [B+\eta'(t')]}
{\cal U}^{\dag}(\tau,0)\sigma_- {\cal U}(\tau,0)\rho_0\right\}\nonumber\\
&=&
\textrm{e}^{i\int_{\tau}^{2\tau}dt' [B+\eta'(t')]}
\textrm{e}^{-i\int_{0}^{\tau}dt' [B+\eta'(t')]}
\rm{Tr}\{\sigma_- \rho_0\}.
\end{eqnarray}
Therefore the double average can be conveniently written as
\begin{equation}
\langle\langle\sigma_+(2\tau)\rangle\rangle = 
\textrm{e}^{-iB\int_0^{2\tau}s(t')dt'}\left\langle
\textrm{e}^{-i\int_0^{2\tau}s(t')\eta'(t')dt'}\right\rangle
\rm{Tr}\left\{\sigma_-\rho_0\right\},
\label{echogen}
\end{equation}
with the introduction of an auxiliary echo function $s(t)$. For Hahn echo   
$s(t)=1$ if $0\leq t<\tau$ and $s(t)=-1$ if $t>\tau$.  
Note that the first term in Eq.~(\ref{echogen}) is exactly equal to
one,  because the Hahn echo is able to completely refocus a constant
magnetic field. 
It is convenient to introduce the noise spectrum in
Eq.~(\ref{echogen}) via $S(t)=\int \textrm{e}^{-i\omega
  t}\tilde{S}(\omega)d\omega$ in order to get the following expression
for the coherence envelope
\begin{equation}
\left|\langle\langle \sigma_{+}(t)\rangle\rangle\right|
=\exp{\left[-\int_{-\infty}^{\infty}d\omega\;\tilde{S}(\omega)
{\cal F}(t,\omega)\right]}.
\label{splus}
\end{equation}
Here we define a filter function that depends on the echo sequence $s(t')$,
\begin{equation}
{\cal F}(t,\omega)=\int_{0}^{t}dt' s(t') \int_{0}^{t'}dt'' s(t'')
\cos{\left[\omega (t'-t'')\right]}.
\end{equation}
For free induction decay [$s(t)\equiv 1$] we have
\begin{equation}
{\cal F}_{\rm{FID}}(t,\omega)=
\frac{1}{2}\frac{\sin^2{\left(\omega t/2\right)}}{\left(\omega/2\right)^2}, 
\label{ffid}
\end{equation}
while for the Hahn echo
\begin{equation}
{\cal F}_{\rm{Hahn}}(2\tau,\omega)=
\frac{1}{2}\frac{\sin^4{\left(\omega \tau/2\right)}}{\left(\omega/4\right)^2}.
\label{fecho}
\end{equation}
Note that ${\cal F}_{\rm{Hahn}}(2\tau,0)=0$. The Hahn echo filters out
terms proportional to $\tilde{S}(0)$ in qubit evolution, this is
equivalent to the well known removal of inhomogeneous broadening by
the echo. Any spin resonance sequence containing instantaneous $\pi/2$ or
$\pi$-pulses (not necessarily equally spaced) can be mapped into an
appropriate echo function $s(t)$. An important example is the class of
Carr-Purcell sequences that can be used to enhance coherence.

\subsubsection{General results for the short time behavior}

We may derive interesting results when the time-dependent
correlation function $S(t)$ is \emph{analytic at $t=0$}. In this case we may
expand around $t=0$ to obtain
\begin{eqnarray} 
S(t)&=&\langle\eta'(t)\eta'(0)\rangle=
\int_{-\infty}^{\infty} d\omega \textrm{e}^{-i\omega
  t}\tilde{S}(\omega)\nonumber\\
&=&\sum_{n=0}^{\infty} \frac{(-1)^n}{(2n)!}M_{2n} t^{2n},
\label{stexpan}
\end{eqnarray}
with the 2n-th moment of the noise spectrum defined as 
\begin{equation}
M_{2n}=\int_{-\infty}^{\infty}
d\omega \tilde{S}(\omega)w^{2n}.
\end{equation}
Hence if $S(t)$ is analytic at $t=0$, we must have $M_{2n}<\infty$
for all $n$, i.e. the noise spectrum has a well defined high frequency
cut-off.  It is important to keep in mind that the assumption of
analyticity at $t=0$ is actually quite restrictive. Physically, only
$M_0<\infty$ is required, so that $S(0)<\infty$ (this is the noise
power or mean square deviation for $\eta'$). Important examples where
$S(t)$ is not analytic at $t=0$ include the Gauss-Markov model
described below [See Eq.~(\ref{expchi})].

If the $t=0$ expansion exists we may immediately
obtain the short time behaviors for the free induction decay and Hahn echo:
\begin{equation}
\langle\langle\sigma_+(t)\rangle\rangle_{\rm{FID}} = 
\textrm{e}^{-\int d\omega
  \tilde{S}(\omega)\frac{1}{2}\frac{\sin^2{(\omega t/2)}}{(\omega/2)^2}}
\approx \textrm{e}^{-\frac{1}{2}M_0 t^2 +\frac{1}{24}M_2 t^4},
\label{genfid}
\end{equation}
\begin{equation}
\langle\langle\sigma_+(2\tau)\rangle\rangle_{\rm{Hahn}} = 
\textrm{e}^{-\int d\omega
  \tilde{S}(\omega)\frac{1}{2}\frac{\sin^4{(\omega \tau/2)}}{(\omega/4)^2}}
\approx \textrm{e}^{-\frac{1}{2}M_2 \tau^4 +\frac{1}{12}M_4 \tau^6}.
\label{genecho}
\end{equation}
The short time behavior described by
Eqs.~(\ref{genfid}),~(\ref{genecho}) is \emph{universal for noise
  spectra possessing a high frequency cut-off}.  Note the striking
difference in time dependence: For free induction decay the coherence
behaves as $\sim\textrm{e}^{-t^2}$, while for a Hahn echo we have
$\sim \textrm{e}^{-\tau^4}$.  This happens because the Hahn echo is
independent of the mean square deviation $M_0=S(0)$.

\subsubsection{Example: The Gauss-Markov model}

The simplest model of Brownian motion assumes a phenomenological
correlation function that decays exponentially in time,
\begin{equation}
S(t)=\Delta^2 \exp{\left(-|t|/\tau_c\right)},
\label{expchi}
\end{equation}
where $\tau_c$ is a correlation time that describes the ``memory'' of
the environmental noise.\footnote{Many authors use the terminology
  ``Markovian dynamics'' to denote evolution without memory, i.e., the
  limit $\tau_c\rightarrow 0$ in Eq.~(\ref{expchi}). This limit can be
  taken by setting $\Delta\rightarrow \infty$ with $\Gamma\equiv
  \Delta^2\tau_c$ held finite. In that case we have $S(t)\rightarrow
  2\Gamma \delta(t)$ resulting in a ``white noise'' spectrum and
  $\langle \sigma_+\rangle\propto \textrm{e}^{-\Gamma t}$.}  This
model is useful e.g. in liquid state NMR in order to calculate the
line-widths of a molecule diffusing across an inhomogeneous magnetic
field. In that case, $\Delta$ becomes the typical field inhomogeneity,
while the ``speed'' for diffusion is of the order of $\Delta/\tau_c$.
The resulting environmental noise spectrum [Fourier transform of
Eq.~(\ref{expchi})] is a Lorentzian, given by
\begin{equation}
\tilde{S}(\omega)=\frac{\Delta^2\tau_c}{\pi}
\frac{1}{\left(\omega\tau_c\right)^2+ 1}.
\label{slorentz}
\end{equation}

We start by discussing free induction decay. Using
Eqs.~(\ref{echogen}) and (\ref{expchi}) with $s(t')=1$,
we get
\begin{equation}
\left|\langle\langle \sigma_+(t)\rangle\rangle\right|_{{\rm FID}} = 
\exp{\left\{
-\Delta^2 \tau_c^2 \left[ t/\tau_c +\left(
\textrm{e}^{-t/\tau_c}-1\right)\right]
\right\}}.
\label{fid}
\end{equation}
For $t\gg \tau_c$, Eq.~(\ref{fid}) leads to
\begin{equation}
|\langle\langle \sigma_+(t)\rangle\rangle|_{{\rm FID}}
\approx \textrm{e}^{-\Delta^2\tau_c t}.
\label{fidmn}
\end{equation}
In this regime, the
correlation function Eq.~(\ref{expchi}) can be approximated by a delta
function, and the decay is a simple exponential signaling that a
Master equation approach is appropriate [Eq.~(\ref{meqn})]. The
coherence time is given by $T^{*}_{2}=1/(\Delta^2\tau_c)$.  Interestingly,
as $\tau_c\rightarrow 0$ with $\Delta$ finite, $T^{*}_{2}\rightarrow\infty$.
This phenomenon is known as motional narrowing, inspired by the motion
of molecules in a field gradient. The faster the molecule is
diffusing, the narrower is its resonance line. Now we look at the low
frequency noise limit, $t\ll \tau_c$.  This leads to 
\begin{equation}
|\langle\langle
\sigma_+(t)\rangle\rangle|_{{\rm FID}} \approx \textrm{e}^{-\frac{1}{2}\Delta^2
  t^2}\equiv \textrm{e}^{-\left(t/T_2^*\right)^2}.
\label{fidlowf}
\end{equation}
In contrast to Eq.~(\ref{fidmn}), the decay differs from a simple
exponential and is independent of $\tau_c$.  This result is equivalent
to an average over an ensemble of qubits at any specific time $t$ (in
other words, the line-width $\Delta/\sqrt{2}$ and dephasing time
$T_2^*=\sqrt{2}/\Delta$ are a consequence of inhomogeneous
broadening). Therefore the coherence decay is completely independent
of the environmental kinetics.  As we shall see below, this decay is
to a large extent reversible by the Hahn echo.

The Hahn echo decay is calculated from
Eqs.~(\ref{splus}),~(\ref{fecho}),~and~(\ref{slorentz}) leading to
\begin{equation}
\langle\langle \sigma_+(2\tau)\rangle\rangle_{{\rm Hahn}} =
\exp{\left\{-\Delta^2 \tau_c^2 \left[ 2\tau/\tau_c -3
+4\textrm{e}^{-\tau/\tau_c}
-\textrm{e}^{-2\tau/\tau_c}
\right]
\right\}}.
\label{hahn}
\end{equation}
For $\tau\gg \tau_c$ we again have motional narrowing,
$\langle\langle \sigma_+(2\tau)\rangle\rangle_{{\rm Hahn}} \approx
\textrm{e}^{-\Delta^2\tau_c 2\tau}$, a result identical to FID
[Eq.~(\ref{fidmn})] if we set $t=2\tau$. This occurs because the noise
trajectories are completely uncorrelated before and after the
$\pi$-pulse, that plays no role in this limit.  For $\tau\ll
\tau_c$ we get
\begin{equation}
\langle\langle \sigma_+(2\tau)\rangle\rangle_{{\rm Hahn}}
\approx\textrm{e}^{-\frac{1}{24}\left(\Delta\tau_c\right)^2
\left(\frac{2\tau}{\tau_c}\right)^3}\approx
\textrm{e}^{-\left(\frac{2\tau}{T_2}\right)^3}.
\label{hahnslow}
\end{equation}
In drastic contrast to free induction decay, Eq.~(\ref{hahnslow})
depends crucially on the kinetic variable $\tau_c$. The time scale
$T_2$ for 1/e decay of Hahn echo\footnote{In the electron spin
  resonance literature the $1/e$ decay time of a Hahn echo is often
  denoted $T_M$. Here follow the spintronics terminology and use $T_2$
  for the $1/e$ decay of Hahn echo, and $T_{2}^{*}$ for $1/e$ decay of
  FID in the low frequency regime.} is considerably longer than
$T_2^*$ when $\Delta\gg \tau_c$.

\subsubsection{A train of Hahn echoes: The Carr-Purcell sequence 
and coherence control}

Consider the sequence
$\pi/2-\left[\tau-\pi-\tau-\rm{echo}\right]_{\rm{repeat}}$. It
consists of the application of a $\pi$-pulse every odd multiple of
$\tau$, with the observation of an echo at even multiples of $\tau$,
i.e. at $t=2n\tau$ for $n$ integer.\footnote{One can make the
  Carr-Purcell sequence robust against pulse errors by alternating the
  phase of the $\pi$-pulses, see e.g. the Carr-Purcell-Meiboom-Gill
  sequence \cite{slichter96}.} In the limit $\tau\ll \tau_c$ the
$n$-th echo envelope can be approximated by a product of $n$ Hahn echoes,
\begin{equation}
\langle\langle \sigma_+(2n\tau)\rangle\rangle_{\rm{CP}}
\approx \langle\langle
\sigma_+(2\tau)\rangle\rangle^{n}_{\rm{Hahn}} \approx
\textrm{e}^{-\frac{2n\tau}{T_2}\left(\frac{2\tau}{T_2}\right)^2}\equiv 
\textrm{e}^{-\frac{2n\tau}{T_2^{\rm{eff}}}},
\label{cpmgdecay}
\end{equation}
with $T_2^{\rm{eff}}\equiv T_2 \left[T_2/(2\tau)\right]^2$. As $\tau$
is decreased below $T_2$ the effective coherence time $T_2^{\rm{eff}}$
increases proportional to $1/\tau^2$. Therefore a train of Hahn echoes
can be used to control decoherence. Rewriting Eq.~(\ref{cpmgdecay})
with $t\equiv 2n\tau$ we get $T_{2}^{\rm{eff}}=(2n)^{2/3}T_2$, showing
that the scaling of the enhanced coherence time with the number of
$\pi$-pulses is sub-linear. The train of $\pi$-pulses spaced by
$\tau\ll \tau_c$ effectively averages out the noise, 
because within $\tau$ much shorter than $\tau_c$ the noise appears to
be time-independent.

\subsubsection{Loss of visibility due to high frequency noise}

In order to understand the role of high frequency noise, 
consider the model Lorentzian noise spectrum peaked 
at frequency $\Omega$ with a broadening given by
$1/\tau_d$,
\begin{equation}
\tilde{S}_{L}(\omega)=\frac{\Delta^2\tau_d}{\pi}
\frac{1}{
\left(\omega-\Omega\right)^2
\tau_{d}^{2}+1}.
\end{equation}
Using Eqs.~(\ref{splus}) and (\ref{ffid}) 
and assuming $\Omega\gg 1/\tau_d$ we get
\begin{equation}
\left|\langle\langle \sigma_+(t)\rangle\rangle\right|_{L} \approx 
\exp{\left[-2\left(\frac{\Delta}{\Omega}\right)^2
  \left(1-\textrm{e}^{-t/\tau_d}\cos{\Omega t}\right)\right]}.
\label{visib}
\end{equation}
Therefore high frequency noise leads to loss of visibility for the
coherence oscillations. The loss of visibility is initially
oscillatory, but decays exponentially to a fixed contrast for $t\gg \tau_d$.
For comparison consider the Gaussian model,
\begin{equation}
\tilde{S}_{G}(\omega)=\frac{\Delta^2}{\sqrt{2\pi\sigma^2}}\exp{\left[
-\frac{(\omega-\Omega)^2}{2\sigma^2}
\right]}.
\end{equation}
For $\Omega\gg \sigma$ we get
\begin{equation}
\left|\langle\langle \sigma_+(t)\rangle\rangle\right|_G \approx 
\exp{\left[-2\left(\frac{\Delta}{\Omega}\right)^2
  \left(1-\textrm{e}^{-\frac{1}{2}\sigma^2 t^2}\cos{\Omega t}\right)\right]}.
\label{visibg}
\end{equation}
Note that the difference between the Gaussian and the Lorentzian
models lies in the time dependence of the approach to a fixed
contrast. Deviations from Lorentzian behavior may be assigned to
non-exponential decays of the coherence envelope. 
Although Eqs.~(\ref{visib}),~(\ref{visibg}) 
were calculated for free induction decay, it
is also a good approximation for Hahn echoes in the limit $\Omega\gg
1/\tau$.

\subsection{Single spin measurement versus ensemble experiments:
  Different coherence times? \label{singlevsmany}}

Recently single shot detection of the spin of a single electron in a
GaAs quantum dot was demonstrated \cite{elzerman04}, and the Hahn echo
decay of the singlet-triplet transition in a double quantum dot was
measured \cite{petta05}.  Also, spin resonance of a single spin center
in the Si/SiO$_2$ interface was detected through time averaged current
fluctuations \cite{xiao04}. These state of the art experiments should
be contrasted with the traditional spin resonance measurements where the
microwave excitation of a sample containing a large number of
localized spins is probed.  Naturally the following question arises:
Are the coherence times extracted from an ensemble measurement any
different from the ones obtained in single spin experiments?

\begin{figure}[t]
\centerline{\includegraphics*[width=.7\textwidth]{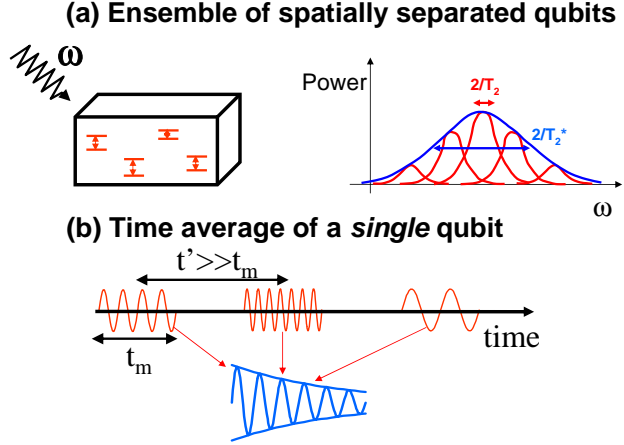}}
\caption[]{(a) A Traditional spin resonance experiment probes the
  coherent evolution of an ensemble of spatially separated spins. The
  spins can be separated into different ``packets'' with similar
  resonance frequencies, each packet with linewidth $2/T_2$. A free
  induction decay measurement is sensitive to the broadened linewidth
  $2/T_{2}^{*}$. A spin echo is needed in order to reveal the
  intrinsic linewidth $2/T_2$. (b) A similar situation applies to
  single spin experiments subject to low frequency noise, because each
  time ensemble may have a different resonance frequency}
\label{fig_single_ensemble} 
\end{figure}

The answer to this question is related to the \emph{ergodicity} of the
environment producing noise, i.e., whether time averages are equal to
ensemble averages. Even when single shot read-out of a quantum degree
of freedom is possible, one must repeat each measurement several
times, in order to obtain good average values for the observables.
For example, measurements of the state of a single spin yields two
possible outcomes and one must time-average an ensemble of identical
qubit evolutions in order to obtain an average value that reflects the
correct outcome probabilities (In Ref.~\cite{elzerman04}, each average
value resulted from $\sim 600$ read-out traces). The presence of phase
fluctuation with correlation time $\tau_c$ smaller than the typical
averaging time implies spin precession with distinct frequencies for
read-out traces separated in time.  This may lead to strong free
induction decay (low $T_{2}^{*}$) in a single shot read-out
measurement, see Fig.~\ref{fig_single_ensemble}. This was indeed observed in
the double-dot experiments of Petta {\it et al} \cite{petta05}.

The free induction decay time $T_{2}^{*}$ in ensemble experiments may
be quite different from single-spin experiments. This is because the
spatial separation of spins adds several new contributions to zero
frequency noise. These include spatially inhomogeneous magnetic
fields, $g$-factors, and strains. However, all these contributions are
removed by a Hahn echo sequence, that filters out environmental
fluctuations with frequencies smaller than $1/\tau$. Therefore we
expect the spin echo decay time $T_2$ for a spatial ensemble and a
single spin to be similar, provided the mechanisms for finite
frequency noise do not vary wildly for spatially separated spins, and
the environment affecting each individual spin is ergodic within the
average time scale (similar reasoning applies for $T_1$, which is
determined by high frequency noise $\omega=\gamma B$).

The Gaussian theory of decoherence described here is appropriate for
ergodic environments. It would be very interesting to explore model
systems both experimentally and theoretically in order to search for
detectable non-ergodic effects in coherent evolution.

\section{Electron spin evolution due to nuclear spins: Isotropic and 
  anisotropic hyperfine interactions, inter-nuclear couplings
  and the secular approximation}

A localized electron spin coupled to a lattice of interacting nuclear
spins provides a suitable model system for the microscopic description
of environmental fluctuations affecting coherent evolution.  Here we
describe a model appropriate for localized electron spins in
semiconductors, and discuss some approximations that can be made
in a moderate magnetic field (typically larger than the inhomogeneous
broadening linewidth, $B>0.1$~T).

\subsection{The electron-nuclear spin Hamiltonian}

The full Hamiltonian for a single electron interacting with N nuclear
spins is given by \cite{slichter96}
\begin{equation}
{\cal H}={\cal H}_{eZ}+{\cal H}_{nZ}+{\cal H}_{en}+{\cal H'}_{en} + 
{\cal H}_{nn}.
\label{htotal}
\end{equation}
Here the Zeeman energies for electron and nuclear spins are respectively
\begin{eqnarray}
{\cal H}_{eZ}&=&\frac{1}{2}\gamma_e B \sigma_z\\
{\cal H}_{nZ}&=&-\gamma_{n}B\sum_i I_{iz},
\end{eqnarray}
where $\bm{\sigma}=(\sigma_x,\sigma_y,\sigma_z)$ is the Pauli matrix
vector representing the electron spin, and $\bm{I}_{i}$ is the
nuclear spin operator for a nucleus located at position $\bm{R}_i$
with respect to the center of the electron wave function. 
For $B=1$~T we have $\gamma_e B\sim 10^{11}$~Hz
while $\gamma_n B\sim 10^8$~Hz. 
The $e-n$ coupling takes place due to isotropic and anisotropic
hyperfine interactions. 
The isotropic hyperfine interaction is given
by 
\begin{equation} 
{\cal H}_{en}=\frac{1}{2}\sum_i A^{\rm{iso}}_{i} \bm{I}_i\cdot \bm{\sigma},
\label{eni}
\end{equation}
with contact hyperfine interaction
\begin{equation}
A^{\rm{iso}}_{i}=\frac{8\pi}{3}\gamma_{e0}\gamma_n\hbar 
\left|\Psi(\bm{R}_i)\right|^2,
\label{ai}
\end{equation}
where $\gamma_{e0}=e/(m_ec)=1.76\times 10^7$~(sG)$^{-1}$ is the
gyromagnetic ratio for a free electron and $\Psi(\bm{r})$ is the
electron's wave function. Typical values of $A^{\rm{iso}}_{i}$ varies
from $A^{\rm{iso}}_{i}\sim 10^7$~Hz for $\bm{R}_i=0$ (at the center of
a donor impurity wave function) to $A^{\rm{iso}}_{i}\sim 0$ for $R_i$
much larger than the impurity Bohr radius. The anisotropic hyperfine
interaction reads
\begin{equation}
{\cal H'}_{en}=\frac{1}{2}\sum_i\bm{I}_i\cdot\bm{A'}_i\cdot \bm{\sigma}
\label{ena}
\end{equation}
with an anisotropic hyperfine tensor given by 
\begin{equation}
(\bm{A}'_i)_{lm}=\gamma_{e0}\gamma_{n}\hbar \int d^3 r 
|\Psi(\bm{r}-\bm{R}_i)|^2\left(\frac{2r}{2r+r_0}\right)
\left(
\frac{3 x_l x_m}{r^5}
-\frac{
\delta_{lm}}{r^3}
\right),\label{aprime}
\end{equation}
where $r_0=e^2/(m_e c^2)$ is the classical electron radius. Note that
Eqs.~(\ref{eni}) and (\ref{ena}) are first order perturbative
corrections in the electron coordinate $\bm{r}$.  
Finally, the nuclear-nuclear
dipolar coupling reads
\begin{equation}
{\cal H}_{nn}=\gamma_n^2\hbar\sum_{i<j} \left[
\frac{\bm{I}_i \cdot\bm{I}_j}{R_{ij}^3} - 
\frac{3\left(\bm{I}_i\cdot\bm{R}_{ij}\right)\left(
\bm{I}_j\cdot \bm{R}_{ij}
\right)}{R_{ij}^5}
\right],
\label{nn}
\end{equation}
where $\bm{R}_{ij}=\bm{R}_i-\bm{R}_j$ is the distance between two
nuclei. The typical energy scale for Eq.~(\ref{nn}) is a few KHz for
nearest-neighbors in a crystal.

The full Hamiltonian Eq.~(\ref{htotal}) is a formidable many-body
problem. It is particularly hard to study because of the lack of
symmetry.  In order to study theoretically the quantum dynamics of an
electron subject to a large number of nuclear spins we need to
truncate Eq.~(\ref{htotal}). Here we discuss some simplifications
appropriate for $B>0.1$~T, a condition typically satisfied in several
experiments.  The first approximation arises when we note that the
electron Zeeman energy is typically $~10^3$ times larger than the
nuclear Zeeman energy. For $B>100$~G the former is much larger than
$A^{\rm{iso}}_i$, therefore the electron spin cannot be ``flipped'' by
the action of the hyperfine interaction. In other words, ``real''
$e-n$ flip-flop transitions get inhibited at these fields (however
virtual transitions induced by second order processes such as ${\cal
  H}_{en}^2$ do produce visible effects as discussed in section
\ref{section_nonsecular}). This consideration allows us to approximate
the isotropic hyperfine interaction to a diagonal form (secular
approximation),
\begin{equation}
{\cal H}_{en}\approx\frac{1}{2}\sigma_z
\sum_i A^{\rm{iso}}_{i} I_{iz}.
\label{diageni}
\end{equation}
The anisotropic hyperfine interaction contains a similar diagonal
contribution in addition to pseudo-secular terms of the form $\sigma_z
I_{i\pm}$. These terms lead to important echo modulations of the order
of $\sim (A'_i/\gamma_n B)^2\sim 0.1-1$ for moderate magnetic fields
($B\sim 0.1-1$~T). To derive these terms, assume $\hat{b}$ as the
direction of the magnetic field and substitute $\bm{\sigma}\rightarrow
\sigma_z \hat{b}$ in Eq.~(\ref{ena}). The result is
\begin{equation}
{\cal H'}_{en}\approx \frac{1}{2}\sigma_z \sum_i \left(
A'_{i\parallel}I_{iz}+
A'_{i\perp}I_{i+}+ A'^{*}_{i\perp} I_{i-}
\right),
\label{ena2}
\end{equation}
where 
\begin{eqnarray}
A'_{i\parallel}&=&\hat{z}\cdot
\bm{A'}_i\cdot \hat{b},\label{apparallel}\\
A'_{i\perp}&=&\frac{1}{2}\left[
\hat{x}\cdot\bm{A'}_{i}\cdot \hat{b}+i\hat{y}\cdot\bm{A'}_i\cdot\hat{b}
\right].\label{apperp}
\end{eqnarray}
For some
lattice sites (closer to the center of the donor impurity) 
$A'_{i\perp}$ is a reasonable fraction of the nuclear spin Zeeman
energy (even for $B\sim 1$~Tesla),
and as a consequence the precession
axis of these nuclear spins will depend on the state of the electron,
producing strong modulations in the nuclear spin echo signal 
\cite{saikin03}.
For the electron spin Eq.~(\ref{ena2}) produces small modulations 
observed at the shortest time scales in the echo decay envelope 
\cite{abe04}. Finally, we can neglect contributions
to Eq.~(\ref{nn}) that do not conserve nuclear spin Zeeman energy, 
\begin{equation}
{\cal H}_{nn}\approx \sum_{i<j}b_{ij} \left(
I_{i+}I_{j-}+I_{i-}I_{j+}-4I_{iz}I_{jz}\right),
\label{nnt}
\end{equation}
with 
\begin{equation}
b_{ij}=-\frac{1}{4}\gamma_n^2\hbar \frac{1-3\cos^{2}{\theta_{ij}}}{R_{ij}^3}.
\end{equation}
Here $\theta_{ij}$ is the angle formed
by the applied $\bm{B}$ field and the vector $\bm{R_{ij}}$ linking the
two nuclear spins $i,j$. This leads to an important orientation dependence
of coherence times.

\subsection{Electron-nuclear spin evolution in the secular approximation}

In the secular approximation [Eqs.~(\ref{diageni})-(\ref{nnt})] the
electron-nuclear spin Hamiltonian is block-diagonal,
\begin{equation}
{\cal H}= {\cal H}_{+} 
\mid\Uparrow\rangle\langle\Uparrow\mid 
+{\cal H}_{-} \mid\Downarrow\rangle\langle\Downarrow\mid,
\label{hupdown}
\end{equation}
where 
$\mid\Uparrow\rangle\langle\Uparrow\mid$ and 
$\mid\Downarrow\rangle\langle\Downarrow\mid$ are projectors in the electron
spin up and down subspaces respectively. Here 
${\cal H}_{\pm}$ contains only nuclear spin operators and is given by
\begin{eqnarray}
{\cal H}_{\pm}&=&{\cal H}_{nn}-\gamma_n B \sum_i I_{iz}\pm
\frac{1}{2}\gamma_e B \nonumber\\
&&\pm \frac{1}{2}\sum_i
\left[A_i I_{iz} +
  A'_{i} I_{i+}+A'^{*}_{i}I_{i-}\right],
\end{eqnarray}
where $A_i\equiv A^{\rm{iso}}_{i}+A'_{i\parallel}$ and $A'_i\equiv
A'_{i\perp}$. Accordingly, the evolution operator becomes ${\cal
  U}(t)= {\cal U}_+(t)\mid\Uparrow\rangle\langle\Uparrow\mid + {\cal
  U}_-(t)\mid\Downarrow\rangle\langle\Downarrow\mid$, with ${\cal
  U}_{\pm}(t)=\textrm{e}^{-it{\cal H}_{\pm}}$.  We can write an
explicit expression for the coherences as a function of the evolution
operators ${\cal U}_{\pm}(t)$ provided the initial density matrix can
be written in product form, $\rho_0=\rho_{0e}\otimes\rho_{0n}$. The
free induction decay is given by
\begin{eqnarray}
\langle\sigma_+(t)\rangle_{\rm{FID}} &=&\rm{Tr}_n\rm{Tr}_e \left\{
{\cal U}^{\dag}\sigma_+{\cal U}\rho_{0e}\otimes\rho_{0n}\right\}\nonumber\\
&=&\rm{Tr}_{n}\left\{{\cal U}^{\dag}_{+}{\cal U}_{-}\rho_{0n}\right\}
\rm{Tr}_{e}\left\{\sigma_+\rho_{0e}\right\},
\label{secularfid}
\end{eqnarray}
where we used the fact that ${\cal U}^{\dag}\sigma_+{\cal U}={\cal
  U}^{\dag}_+{\cal U}_-
  \mid\Uparrow\rangle\langle\Downarrow\mid$, and
  $\sigma_+=\mid\Uparrow\rangle\langle\Downarrow\mid$. For the Hahn
  echo, we use ${\cal U}(\tau)\sigma_x{\cal U}(\tau)$ as our evolution
  operator, to get
\begin{equation}
\langle \sigma_+(2\tau)\rangle_{\rm{Hahn}} = \rm{Tr}_n \left\{{\cal
    U}^{\dag}_{-}{\cal U}^{\dag}_{+}{\cal U}_{-}{\cal U}_{+}\rho_{0n}\right\}
\rm{Tr}_e \left\{\sigma_-\rho_{0e}\right\}.
\label{secularhahn}
\end{equation}
Eqs.~(\ref{secularfid}) and (\ref{secularhahn}) are exact in the
secular approximation, and make explicit the dependence of the
electron's coherence envelope in the nuclear spin Hamiltonian
evolution. 

\subsubsection{Inhomogeneous broadening due to the isotropic 
hyperfine interaction}

The diagonal model 
\begin{equation}
{\cal H}_{\pm}=-\gamma_nB \sum_i I_{iz}\pm \frac{1}{2}\gamma_eB \pm
\frac{1}{2}\sum_i A_i I_{iz}
\end{equation}
is easily solved exactly for nuclear spins initially in a product
state. Assume the electron spin is pointing in the $+y$ direction, and
the nuclear spin states are distributed randomly, each nuclei with equal
probability of pointing up or down.
The free induction decay amplitude becomes
\begin{eqnarray}
\langle \sigma_+(t)\rangle_{\rm{FID}}&=&\frac{i}{2}
\rm{Tr}_n \left\{{\cal U}^{\dag}_{+}{\cal U}_{-}
\right\}\nonumber\\&=&
\frac{i}{2}\textrm{e}^{i\gamma_e Bt} 
\prod_j \frac{1}{2}\left[\textrm{e}^{\frac{i}{2}A_j t}+
\textrm{e}^{-i\frac{i}{2}A_j t}\right]\nonumber\\
&=&\frac{i}{2}\textrm{e}^{i\gamma_e Bt} 
\sum_{\xi_1=\pm1,\ldots}
\frac{1}{2^N}\textrm{e}^{\frac{i}{2}t\sum_j
A_j \xi_j}\nonumber\\
&\approx&\frac{i}{2}\textrm{e}^{i\gamma_e Bt} 
\textrm{e}^{-\frac{1}{8}t^2\sum_j A_j^2},
\label{inhomobroad}
\end{eqnarray}
where in the last line we assumed $N\rightarrow\infty$ with each
individual $A_i\rightarrow 0$ so that the hyperfine field can be
approximated by a continuous Gaussian distribution. The free induction
decay rate or inhomogeneously broadened linewidth is given by
$\frac{1}{T_2^*}\sim A_{\rm{rms}}=\sqrt{\sum_jA_j^2}$. This fast decay
rate should be compared to the Hahn echo: From Eq.~(\ref{secularhahn})
we see that $\langle \sigma_+(2\tau)\rangle_{\rm{Hahn}}=-i/2$, the
Hahn echo never decays.  In fact, from Eq.~(\ref{secularhahn}) we can
easily prove that the class of Hamiltonians of the form Eq.~(\ref{hupdown})
satisfying $\left[{\cal H}_+,{\cal H}_-\right]=0$ have time
independent Hahn echoes given by $\rm{Tr}\{\sigma_-\rho_{0e}\}$
\cite{shenvi05b}.

%


\subsection{Beyond the secular approximation: Nuclear-nuclear
  interactions mediated by the electron spin hyperfine interaction}
\label{section_nonsecular}

In the sections above we showed that the secular approximation allows
us to decouple electron spin dynamics from nuclear spin dynamics
completely.  This approximation clearly does not hold at low magnetic
fields, and the problem becomes considerably more complicated.  The
study of electron spin evolution subject to the full isotropic
hyperfine interaction has attracted a great deal of attention lately
\cite{merkulov02,khaetskii02,shenvi05b,deng06}, particularly
because of a series of free induction decay experiments probing
electron spin dynamics in quantum dots in the low magnetic field
regime \cite{gammon01}.  In the author's opinion the most successful
theoretical approach so far in the description of these experiments is
to treat the collective nuclear spin field classically by taking
averages over its direction and magnitude \cite{merkulov02}. Here we
shall not discuss the interesting effects occurring at low fields. Instead,
we will focus on the following question: What is the threshold field
$B_{\rm th}$ for the secular approximation to hold? For intermediate
$B>B_{\rm th}$ (not satisfying $B\gg B_{\rm th}$), how can the
non-secular terms be incorporated in a block diagonal Hamiltonian of
the form Eq.~(\ref{hupdown})?

In order to answer these questions, let's consider the Hamiltonian
\begin{equation}
{\cal H}={\cal H}_0 +{\cal V},
\label{htnonsecular}
\end{equation} 
\begin{equation}
{\cal H}_0=\frac{1}{2}(\gamma_e+\gamma_n)B\sigma_z
+\frac{1}{2}\sigma_z
\sum_j A_j I_{jz},
\end{equation}
\begin{equation}
{\cal V}=\frac{1}{2}\sum_j A_j\left(\sigma_+I_{j-}+\sigma_-I_{j+}\right),
\label{vnonsec}
\end{equation}
where ${\cal H}_0$ and ${\cal V}$ denote the secular and non-secular
contributions respectively. Here we remove the nuclear Zeeman energy
by transforming to the rotating frame precessing at $\gamma_n B$. From
Eq.~(\ref{vnonsec}) we may be tempted to assume that flip-flop
processes involving an electron and a nuclear spin (e.g.,
$\Uparrow\downarrow\rightarrow\Downarrow\uparrow$) are forbidden by
energy conservation at fields $\gamma_e B\gg A_i$. However, the
situation is much more complex because higher order ``virtual''
processes such as ${\cal V}^2=A_iA_j I_{i+}I_{j-}\sigma_+\sigma_-
+\cdots$ preserve the electron spin polarization and hence may have a
small energy cost (of the order of $A_i-A_j$ for $A_i\sim A_j$). As we
show below, these processes actually lead to a long-range effective
coupling between nuclear spins, similar to the RKKY interaction
between nuclear spins in a metal. We will show this using the original
self-consistent approach of Shenvi {\it et al}. 

Let $\ket{\psi^+}$ be a ``+" eigenstate of the Hamiltonian
Eq.~(\ref{htnonsecular}), i.e. $\ket{\psi^+}$ has primarily electron
spin-up character.  Without loss of generality, $\ket{\psi^+}$ can be
written as
\begin{equation}
\ket{\psi^+} = \ket{\Uparrow,\psi_\Uparrow^+} + 
\ket{\Downarrow,\psi_\Downarrow^+}.
\end{equation}
Because the perturbation ${\cal V}$ flips the polarization of the electron, 
the 
action of ${\cal H}$ on the electron spin-up and electron spin-down 
subspaces yields the two simultaneous equations,
\begin{eqnarray}\label{Equation::HPsiPlus}
{\cal H}_0 \ket{\Uparrow,\psi_\Uparrow^+} + {\cal V} 
\ket{\Downarrow,\psi_\Downarrow^+} 
&=& E_+ \ket{\Uparrow,\psi_\Uparrow^+},\\
\label{Equation::HPsiMinus}
{\cal H}_0 \ket{\Downarrow,\psi_\Downarrow^+} + {\cal V} 
\ket{\Uparrow,\psi_\Uparrow^+} 
&=& E_+ \ket{\Downarrow,\psi_\Downarrow^+}.
\end{eqnarray}
\refeqn{HPsiMinus} can be solved for 
$\ket{\Downarrow,\psi_\Downarrow^+}$ 
and the resulting expression inserted into \refeqn{HPsiPlus} yields
\begin{equation} \label{Equation::WignerPsiPlus}
{\cal H}_0 \ket{\Uparrow,\psi_\Uparrow^+} + {\cal V} \frac{1}{E_+ -
  {\cal H}_0} {\cal V} 
\ket{\Uparrow,\psi_\Uparrow^+} = 
E_+ \ket{\Uparrow, \psi_\Uparrow^+}.
\end{equation}
In the presence of an energy gap between the spin-up and spin-down
states (this is certainly true at high magnetic fields satisfying $B>
\sum A_j/\gamma_e$), the operator $1/(E_+-{\cal H}_0)$ is
always well-defined \cite{shenvi05a}.  Because the left-hand side of
\refeqn{WignerPsiPlus} depends on $E_+$, it is not a true
Schr\"{o}dinger equation; to obtain $E_+$ exactly,
\refeqn{WignerPsiPlus} must be solved self-consistently.  However, if
we use $E_+ \approx (\gamma_e + \gamma_n) B / 2$, then we can
obtain an effective Hamiltonian from \refeqn{WignerPsiPlus}.  The
effective Hamiltonian in the electron spin-up subspace is
\begin{eqnarray} \label{Equation::Heffp}
{\cal H}_{\mathrm{eff}}^+ &=& {\cal H}_0 + {\cal V}_{\mathrm{eff}}^+ \\
\label{Equation::Veffp}
{\cal V}_{\mathrm{eff}}^+ &=& \frac{1}{4}\sum_{j,k}{A_j A_k I_{j-} 
\frac{1}{(\gamma_e+\gamma_n)B+\frac{1}{2}\sum_j{A_j I_{jz}}}I_{k+}}.
\end{eqnarray}
We obtain a similar, but not identical, 
effective Hamiltonian for the spin-down 
subspace (note the transposition of the $I_-$ and $I_+$ operators),
\begin{eqnarray} \label{Equation::Heffm}
{\cal H}_{\mathrm{eff}}^- &=& {\cal H}_0 + {\cal V}_{\mathrm{eff}}^- \\
\label{Equation::Veffm}
{\cal V}_{\mathrm{eff}}^- &=& -\frac{1}{4}\sum_{j,k}{A_j A_k I_{j+} 
\frac{1}{(\gamma_e-\gamma_n)B+\frac{1}{2}\sum_j{A_j I_{jz}}}I_{k-}}.
\end{eqnarray}
\refeqs{Veffp}{Veffm} show that the overall coupling between nuclei
does indeed decrease at high fields, because the operator $1/(E-{\cal
  H}_0)$ scales approximately as $1/B$.  However, the energy cost for
flip-flopping two nuclei $j$ and $k$ is proportional to $A_j - A_k$.
Thus, if $A_j$ and $A_k$ are close in value, the nuclei can flip-flop
even at high fields.
Eqs.~(\ref{Equation::Veffp}),~(\ref{Equation::Veffm}) were later
derived using an alternative canonical transformation approach
\cite{yao05}.

We may expand \refeqs{Veffp}{Veffm} in powers of $\sum_j A_j
I_{jz}/(\gamma_e B)$, so that for the unpolarized case we have
approximately
\begin{equation}
{\cal V}_{\rm{eff}}\approx\frac{1}{2}\sigma_z \sum_{j,k}
\frac{A_jA_k}{2(\gamma_e+\gamma_n) B}I_{j+}I_{k-}.
\label{veffapprox}
\end{equation}
This effective Hamiltonian is of the secular type
Eq.~(\ref{hupdown}), and satisfies the symmetry condition $\left[{\cal
    H}_+,{\cal H}_-\right]=0$. Therefore a Hahn echo is able to
refocus this interaction completely. 

Shenvi {\it et al} performed exact numerical calculations of electron
spin echo dynamics in clusters of $N=13$ nuclear spins
\cite{shenvi05b}. The Hahn echo envelope was found to decay fast to a
loss of contrast given by
\begin{equation} 
\left|\langle \sigma_+(2\tau)\rangle\right|\approx
\frac{1}{2}-\frac{\sum_j A_j^2}{\left[(\gamma_e+\gamma_n) B\right]^2}.
\end{equation}
This shows that the threshold field for neglecting the non-secular
isotropic hyperfine interaction in the Hahn echo is given by the
inhomogeneously broadened linewidth, $B_{\rm th}=\sqrt{\sum_j
  A_j^2}/\gamma_e\sim 10-100$~G (for a donor impurity in silicon,
$B_{\rm th}$ is $\approx 1$~G for natural samples and $\approx 10$~G
for $^{29}$Si enriched samples). 

Recently, Yao {\it et al} \cite{yao05} and Deng {\it et al.}
\cite{deng06} showed that the electron-mediated inter-nuclear coupling
may be observed as a magnetic field dependence of the free induction
decay time in small quantum dots. There is currently an interesting
debate on the correct form of the time dependence for FID decay.  Yao
{\it et al.} derived the FID decay from Eq.~(\ref{veffapprox}) and
obtained $\langle \sigma_+(t)\rangle\sim \textrm{e}^{-t^2}$, while
Deng {\it et al.} carried out a full many-body calculation to argue
that FID scales as a power law according to $\langle
\sigma_+(t)\rangle \sim 1/t^2$.


\section{Microscopic calculation of the nuclear spin noise spectrum 
and electron spin decoherence}
\label{section_nuclear}

In this section we discuss low frequency noise due to interacting
nuclear spins. For simplicity, we assume only isotropic hyperfine
interaction. The inclusion of anisotropic hyperfine interaction is
considerably more complicated, but can be seen to lead to high
frequency noise and echo modulations (at frequencies close to
$\gamma_n B\sim 10^8$~s$^{-1}$ per Tesla).  In the next section we
provide explicit numerical calculations for the case of
a phosphorus impurity in silicon and compare our results to
experiments.

\subsection{Nuclear spin noise}

Using the approximations Eqs.~(\ref{diageni}) and (\ref{nnt}) we can
write the electron-nuclear Hamiltonian in a form similar to
Eq.~(\ref{ht}). In the electron spin Hilbert space, we assume an
effective time-dependent Hamiltonian of the form
\begin{equation}
{\cal H}_{\rm{eff}}^{e}=\frac{1}{2}\gamma_e B\sigma_z + 
\frac{1}{2}\sigma_z \sum_i A_i I_{iz} (t),
\label{eeff}
\end{equation}
with $A_i\equiv A^{\rm{iso}}_{i}$.
Nuclear spin noise is in turn determined by the effective
Hamiltonian
\begin{equation}
{\cal H}_{\rm{eff}}^{n}=\sum_{i<j} {\cal H}_{ij}^{n},
\end{equation}
\begin{eqnarray}
{\cal H}_{ij}&=&\gamma_n B \left(I_{iz}+I_{jz}\right)+ b_{ij}\left(
I_{i+}I_{j-}+I_{i-}I_{j+}-4I_{iz}I_{jz}\right) \nonumber\\
&&+
\frac{1}{2}\left(
A_i I_{iz}+A_j I_{jz}\right),
\label{hnij}
\end{eqnarray}
where we decoupled the electron spin from the nuclear spins by
assuming the nuclear spin wave function evolves in the electron spin
up subspace ($\sigma_z \rightarrow 1$). An equally valid choice is to
assume $\sigma_z\rightarrow -1$. It turns out that this choice does
not matter within the pair approximation described below. We will
check this by noting that the final answer is unchanged under the
operation $A_i\rightarrow -A_i$ for all $i$.\footnote{As we shall see
  below, this approximation leads to identical results as the lowest
  order cluster expansion developed in \cite{witzel05}.  However,
  interesting interference effects arise when this approximation is
  not valid.  The cluster expansion method beyond lowest order
  \cite{witzel06} takes account of the full electron-nuclear
  evolution, therefore it can be used to study these effects.}

\begin{figure}[t]
\centerline{\includegraphics*[width=.7\textwidth]{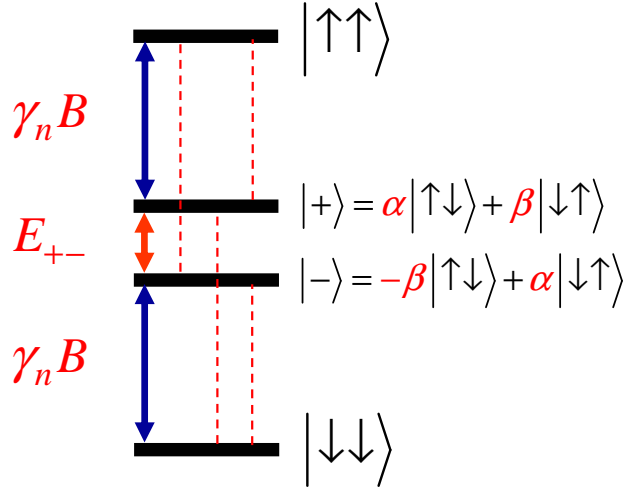}}
\caption[]{Energy levels for two nuclear spins coupled through the
  dipolar interaction. The flip-flop mechanism corresponds to
  transitions between the states $|+\rangle$ and $|-\rangle$, which
  are admixtures of $\mid \uparrow\downarrow\rangle$ and
  $\mid\downarrow\uparrow$ states. The anisotropic hyperfine
  interaction couples states differing by $\sim \gamma_nB$ in energy.}
\label{fig_levels} 
\end{figure}

For now we assume the nuclear spins are unpolarized
($T=\infty$) so that $\langle \sum_i A_i I_{iz}\rangle=0$. This
approximation will be relaxed below.  The time dependent correlation
function for nuclear spins is given by
\begin{eqnarray}
S(t)&=&\left\langle \sum_i A_i I_{iz}(t)
\sum_{j} A_j I_{jz}(0) \right\rangle\nonumber\\
&=& \sum_i A_i^2 \langle I_{iz}(t)I_{iz}(0)\rangle
+ \sum_{i, j(\neq i)} A_iA_j \langle I_{iz}(t)I_{jz}(0)\rangle.
\label{st}
\end{eqnarray}
We now invoke a ``pair approximation'' by assuming
\begin{eqnarray}
\langle I_{iz}(t)I_{iz}(0)\rangle \approx \sum_{j(\neq i)}
\langle I_{iz}(t)I_{iz}(0)\rangle_{ij},\label{pair1}\\
\langle I_{iz}(t)I_{jz}(0)\rangle \approx 
\langle I_{iz}(t)I_{jz}(0)\rangle_{ij},\label{pair2}
\end{eqnarray}
where $\langle \cdot \rangle_{ij}$ denotes a thermal average
restricted to the $ij$ Hilbert space. The operator $I_{iz}(t)$ is in
the Heisenberg representation defined by the two particle Hamiltonian
Eq.~(\ref{hnij}). Plugging Eqs.~(\ref{pair1})~and~(\ref{pair2}) in
Eq.~(\ref{st}) and reordering terms we get
\begin{equation}
S(t)\approx \sum_{i<j} \langle \hat{\eta}_{ij}(t)
\hat{\eta}_{ij}(0)\rangle_{ij},
\label{sthateta}
\end{equation}
with
\begin{equation}
\hat{\eta}_{ij}=A_i I_{iz}+A_j I_{jz}.
\label{hatetaij}
\end{equation}
The same derivation could be given for finite temperature, when the
thermal average of the hyperfine field is non-zero. The result is
identical to Eq.~(\ref{sthateta}) except for the substitution
$\hat{\eta}_{ij}\rightarrow\delta\hat{\eta}_{ij}
=\hat{\eta}_{ij}-\langle\hat{\eta}_{ij}\rangle$.  Using the definition
of the noise spectrum [Eq.~(\ref{noise})] and expanding the correlator
in the energy eigenstates of the pair Hamiltonian Eq.~(\ref{hnij}) we
get
\begin{equation}
\tilde{S}_{ij}(\omega)=\sum_{\alpha,\beta}p_\alpha
\left|\langle \alpha|
  \delta\hat{\eta}_{ij}|\beta\rangle \right|^2
\delta\left(\omega-E_{\beta\alpha}\right),
\label{sijtilde}
\end{equation}
with $E_{\beta\alpha}=E_{\beta}-E_{\alpha}$ the energy difference
between the energy eigenstates $|\alpha\rangle, |\beta\rangle$, and
$p_{\alpha}$ the (thermal) occupation of state $\alpha$. Therefore the
noise spectrum is a sum over all possible transition frequencies
induced by the operator $\hat{\eta}_{ij}$.  For nuclear spin $1/2$ the
$ij$ Hamiltonian has the following eigenenergies and eigenstates (See
Fig.~\ref{fig_levels})
\begin{eqnarray}
E_{\uparrow\uparrow}&=&\gamma_n B -b_{ij}
 +a_{ij},\\
E_{+}&=&b_{ij}+\sqrt{b_{ij}^2+\Delta_{ij}^2},\label{eeplus}\\
E_{-}&=&b_{ij}-\sqrt{b_{ij}^2+\Delta_{ij}^2},\label{eeminus}\\
E_{\downarrow\downarrow}&=&-\gamma_n B -b_{ij}-a_{ij},
\end{eqnarray}
\begin{eqnarray}
|+\rangle &=& 
\cos{\frac{\theta}{2}}\mid\uparrow\downarrow\rangle
 +\sin{\frac{\theta}{2}}\mid\downarrow\uparrow\rangle,
\label{statep}\\
|-\rangle &=& 
-\sin{\frac{\theta}{2}}\mid\uparrow\downarrow\rangle
 +\cos{\frac{\theta}{2}}\mid\downarrow\uparrow\rangle,\label{statem}
\end{eqnarray}
with 
\begin{eqnarray}
a_{ij}&=&\frac{1}{4}\left(A_i+A_j\right),\\
\Delta_{ij}&=&\frac{1}{4}\left(A_i-A_j\right),\\
\cos{\theta}&=&\frac{\Delta_{ij}}{\sqrt{b_{ij}^2+\Delta_{ij}^2}},\\
\sin{\theta}&=&\frac{b_{ij}}{\sqrt{b_{ij}^2+\Delta_{ij}^2}}.
\end{eqnarray}
Using Eqs.~(\ref{statep}),~(\ref{statem}) the transition matrix
element is easily found to be 
\begin{equation}
\langle -|\hat{\eta}_{ij}|+\rangle =
-4\Delta_{ij}\sin{\frac{\theta}{2}}\cos{\frac{\theta}{2}}=
-2\Delta_{ij}\sin{\theta}.
\end{equation}
The transition frequency is simply the difference between
Eqs.~(\ref{eeplus}) and (\ref{eeminus}),
\begin{equation}
E_{+-}=2\sqrt{b_{ij}^2+\Delta_{ij}^2}.
\label{epminus}
\end{equation}
The resulting noise spectrum is therefore
\begin{equation}
\tilde{S}_{ij}(\omega)=\left(A^{{\rm rms}}_{ij}\right)^{2}\delta(\omega)+
4\frac{
b_{ij}^2
\Delta_{ij}^2
}{b_{ij}^2+\Delta_{ij}^2}
\left[p_+ \delta(\omega+E_{+-})+p_- \delta(\omega-E_{+-})
\right],
\label{sijexp}
\end{equation}
with a static contribution given by
\begin{eqnarray}
\left(A^{{\rm rms}}_{ij}\right)^2&=&4\left[\left(p_{\uparrow\uparrow}
+p_{\downarrow\downarrow}\right)a_{ij}^{2}+\left(p_{+}+p_-\right)
\frac{\Delta_{ij}^{4}}{b_{ij}^{2}
+\Delta_{ij}^{2}}\right]\nonumber\\
&&-4\left[\left(p_{\uparrow\uparrow}-p_{\downarrow\downarrow}\right)a_{ij}
+\left(p_+-p_-\right) \frac{\Delta_{ij}^{2}}{\sqrt{b_{ij}^{2}+\Delta_{ij}^{2}}}
\right]^2.
\label{staticcon}
\end{eqnarray}

The free induction decay due to this noise spectrum can be easily
calculated using Eq.~(\ref{splus}) and the filter function
Eq.~(\ref{ffid}),
\begin{eqnarray}
\langle\langle\sigma_+(t)\rangle\rangle &=& 
\exp{\left[-\frac{1}{2}\sum_{i<j}
\left(A^{{\rm rms}}_{ij}\right)^2t^2\right]}\nonumber\\
&\times&\exp{\left[-2\sum_{i<j}\left(p_++p_-\right)\frac{b_{ij}^2
\Delta_{ij}^{2}}{b_{ij}^2+\Delta_{ij}^2}t^2
{\rm sinc}^2\left(\sqrt{b_{ij}^2+\Delta_{ij}^2}t\right) 
\right]},
\label{fidij}
\end{eqnarray}
where $\rm{sinc}(x)=\sin{(x)}/x$. As expected, the FID decay is usually
dominated by the zero-frequency noise amplitue $A^{{\rm rms}}_{ij}$
[Eq.~(\ref{staticcon})]. The dipolar-induced decay may be visible
provided $A^{{\rm rms}}_{ij}$ is much smaller than the finite
frequency noise amplitudes. For example, this is the case if the
nuclear spins are polarized, since we have $A^{{\rm rms}}_{ij}=0$
exactly when $p_{\downarrow\downarrow}=1$ in Eq.~(\ref{staticcon}).

The Hahn echo decay envelope is derived after integration with
the filter function Eq.~(\ref{fecho}),
\begin{equation}
\langle\langle\sigma_+(2\tau)\rangle\rangle =
\exp{\left[-8\sum_{i<j}
\left(p_+ +p_-\right)
b_{ij}^2
\Delta_{ij}^2
\tau^4 {\rm
  sinc}^4\left(\sqrt{b_{ij}^2+\Delta_{ij}^2}\tau\right)
\right]}.
\label{expij}
\end{equation}
At $T\rightarrow \infty$ ($p_{+}=p_-=1/4$) Eq.~(\ref{expij}) is
identical to the echo decay obtained by two completely different
methods, viz.  the lowest order cluster expansion [Eq.~(20) in
Ref.~\cite{witzel05}] and the quasiparticle excitation model [Eq.~(18)
in Ref.~\cite{yao05}]. As expected, Eq.~(\ref{expij}) is independent
of zero-frequency noise, and is exactly equal to $1$ when either
$b_{ij}=0$ or $\Delta_{ij}=0$, or when the nuclear spins are polarized
($p_+=p_{-}=0$).

By expanding the exponent in Eqs.~(\ref{fidij}) and (\ref{expij}) in
powers of time, we find that only even powers are present. The
short time behavior for FID is $\sim \textrm{e}^{-t^2}$, while for
Hahn echo $\sim \textrm{e}^{-\tau^4}$. This short time approximation
is valid for times much shorter than the inverse cut-off frequency of
the noise spectrum obtained after summing over all pairs $i,j$.

\subsection{Mean field theory of noise broadening: quasiparticle
  lifetimes\label{section_broadening}}

\begin{figure}[t]
\centerline{\includegraphics*[width=.7\textwidth]{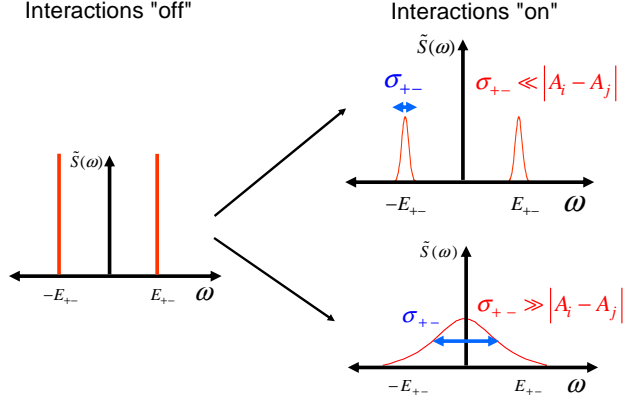}}
\caption[]{An isolated pair of nuclear spins with infinite lifetime
  will produce a sharp peak in the noise spectrum. The role of the
  many-body interactions with other nuclear spins is to broaden this
  peak and smooth out the noise spectrum for the collective nuclear
  spin excitations. Here we calculate the line broadening for each
  pair flip-flop transition using a procedure similar to van Vleck's
  method of moments.}
\label{fig_broadening} 
\end{figure}

We showed that the noise spectrum due to flip-flop transitions in the
Hilbert space formed by two nuclear spins $i,j$ is a linear
combination of delta functions. We may extend this pair approximation
to clusters larger than two, and the number of delta functions will
grow exponentially with cluster size [This can be done by
sistematically increasing the size of the Hilbert space beyond a single
pair $i,j$ in Eqs.~(\ref{pair1})~and~(\ref{pair2})].  These delta
functions can be interpreted as well defined nuclear spin excitations
with infinite lifetime.\footnote{In Ref.~\cite{yao05} Yao {\it et al}
  derive a similar quasiparticle picture via direct calculation of the
  time dependent correlation function for the electron spin.  However,
  the authors did not calculate the quasiparticle relaxation times.
  The noise spectrum is a natural starting point for developing a
  theory for quasiparticle energy broadening, as we show here.} With
this interpretation in mind, it is natural to expect that the
many-body interactions of the pair $i,j$ with other nuclear spins will
limit the lifetime of the flip-flop excitations, i.e., the delta
functions will be broadened, see Fig.~\ref{fig_broadening}.

We can add broadening to the delta functions in a mean field fashion
by using the method of moments, which is applicable at infinite
temperature (no nuclear spin polarization, i.e. $k_BT\gg 
\gamma_n B$). In this limit, the noise spectrum is written as
\begin{equation}
\tilde{S}_{ij}(\omega)=\sum_{\alpha,\beta}\frac{1}{2^N}
|\langle\alpha|\hat{\eta}_{ij}|\beta\rangle|^2 \delta(\omega-E_{\beta\alpha}),
\end{equation}
where here $\alpha,\beta$ denote exact many-body eigenstates of the
system of $N$-coupled nuclear spins.  The $n$-th moment
$\int\omega^n\tilde{S}(\omega)d\omega$ can be calculated exactly using
the invariance of the trace.\footnote{A similar method was used in the
  semiclassical theory of spectral diffusion in order to calculate the
  flip-flop rates for pairs of nuclear spins \cite{desousa03b}.}
Consider the zeroth-moment,
\begin{eqnarray}
\int_{-\infty}^{\infty} \tilde{S}_{ij}(\omega)d\omega &=&
\frac{1}{2^N}\sum_{\alpha,\beta}
\langle\alpha|\hat{\eta}_{ij}|\beta\rangle
\langle\beta|\hat{\eta}_{ij}|\alpha\rangle\nonumber\\
&&=\frac{1}{2^N}\rm{Tr}\left\{
\hat{\eta}_{ij}^{2}
\right\}=\frac{1}{4}\left(A_{i}^{2}+A_{j}^{2}\right).
\label{zeromoment}
\end{eqnarray}
Accordingly, the second moment is given by 
\begin{eqnarray}
\int_{-\infty}^{\infty} \omega^2\tilde{S}_{ij}(\omega)d\omega &=&
-\frac{1}{2^N}\rm{Tr}\left\{
\left[{\cal H}_{eff}^{n},\hat{\eta}_{ij}\right]^2
\right\}\nonumber\\
&=& \frac{1}{2}A_{i}^2\sum_{k\neq i} b_{ik}^2 - b_{ij}^2 A_i A_j 
+\frac{1}{2}A_j^2\sum_{k\neq j}b_{jk}^2.
\label{secondmoment}
\end{eqnarray}
Note that Eqs.~(\ref{zeromoment}) and (\ref{secondmoment}) are exact
at infinite temperature. 

The mean field approximation employed here assumes each delta function
in the noise spectrum can be approximated by a Gaussian function
normalized to one.\footnote{We can verify this assumption by
  calculating the skewness (fourth moment divided by three times the
  second moment squared). For a perfect Gaussian the skewness is
  exactly one. We carried out this calculation and showed
  that for $\Delta_{ij}<\sim b_{ij}$ the skewness is very close to
  one. On the other hand for $\Delta_{ij}\gg b_{ij}$ the skewness
  becomes large, and a better approximation is a Lorentzian with a
  cut-off at the wings. Nevertheless, by inspecting Eq.~(\ref{expij})
  we note that nuclear spin pairs with $\Delta_{ij}\gg b_{ij}$ give a
  much weaker contribution to echo decay than pairs with
  $\Delta_{ij}\sim b_{ij}$.  Therefore this Gaussian fit is precisely
  valid for most important pairs.  As discussed in
  Eqs.~(\ref{visib}),~(\ref{visibg}) the difference between a Gaussian
  and a Lorentzian fit lies in the time dependence of the decay of
  coherence modulations; This is $\sim \textrm{e}^{-\sigma^2t^2/2}$
  for a Gaussian and $\sim \textrm{e}^{-t/\tau_d}$ for a Lorentzian.}
The noise spectrum becomes
\begin{equation}
\tilde{S}_{ij}(\omega)\approx\sum_{\alpha,\beta}
\frac{1}{4}|\langle\alpha|\hat{\eta}_{ij}|\beta\rangle|^2 
\frac{1}{\sqrt{2\pi \sigma_{\alpha\beta}^{2}}}\exp{\left[
-\frac{(\omega-E_{\alpha\beta})^2}{2\sigma_{\alpha\beta}^{2}}
\right]},
\label{gaussiannoise}
\end{equation}
and the second moment is 
\begin{equation}
\int_{-\infty}^{\infty} \omega^2 \tilde{S}_{ij}(\omega)d\omega
\approx \sum_{\alpha,\beta}
\frac{1}{4}|\langle\alpha|\hat{\eta}_{ij}|\beta\rangle|^2 
\left(\sigma_{\alpha\beta}^{2}+E_{\alpha\beta}^{2}\right).
\label{gaussianmoment}
\end{equation}
We now calculate the broadenings $\sigma_{\alpha\beta}$ by equating
Eq.~(\ref{gaussianmoment}) with Eq.~(\ref{secondmoment}).  This
procedure can be carried out exactly, since the
noise spectrum has two identical peaks (Fig.~\ref{fig_broadening}) at
frequencies $\pm E_{+-}$. The broadening is found to be
\begin{equation}
\sigma_{+-}^{2}=\frac{b_{ij}^{2}+\Delta_{ij}^{2}}{4\Delta_{ij}^{2}b_{ij}^{2}}
\sum_{k\neq i,j}\left(
b_{ik}^{2}A_{i}^{2}+b_{jk}^{2} A_{j}^{2}
\right).
\label{sigpm}
\end{equation}
When $\Delta_{ij}< \sim b_{ij}$, 
the broadening becomes of the same order of magnitude as van Vleck's
second moment for the dipolar interaction (equal to
$9\sum_{k}b_{ik}^{2}$ \cite{slichter96}).
For $\Delta_{ij}\gg b_{ij}$ we have $\sigma_{+-}\sim
\Delta_{ij}$, and $\sigma_{+-}/E_{+-}\sim 1$. This type of excitation
is of high frequency and short lifetime, showing fast decay to a small
loss of contrast as described by Eqs.~(\ref{visib}), (\ref{visibg}).
The broadenings describe the diffusion of localized nuclear spin
excitations (deviations from thermal equilibrium) over length scales
greater than the pair distance. 

By adding broadenings to the delta functions in Eq.~(\ref{sijexp}) we
are able to plot a smooth noise spectrum, and study the relative
contributions of a large number of nuclear spins as a function of a
continuous frequency. The modified Eq.~(\ref{sijexp}) summed over all
nuclear pair contributions reads
\begin{eqnarray}
\tilde{S}(\omega)&=&
4\sum_{i<j}\frac{b_{ij}^2
\Delta_{ij}^2}{b_{ij}^2+\Delta_{ij}^2}
\left\{
p_+\frac{\textrm{e}^{-\frac{(\omega-E_{+-})^2}{2\sigma^{2}_{+-}}}}
{\sqrt{2\pi\sigma^{2}_{+-}}}+
p_-\frac{\textrm{e}^{-\frac{(\omega+E_{+-})^2}{2\sigma^{2}_{+-}}}}
{\sqrt{2\pi\sigma^{2}_{+-}}}
\right\}\nonumber\\
&&+\sum_{i<j}\left(A^{{\rm rms}}_{ij}\right)^{2}\delta(\omega).
\label{sijtotal}
\end{eqnarray}
For studies of echo decay we may drop the delta function contribution
at zero frequency. Note that the first part of Eq.~(\ref{sijtotal})
gives an additional zero frequency contribution, that is the limit
$\omega\rightarrow 0$ of the broadened spectrum.

\section{Electron spin echo decay of a phosphorus impurity in silicon:
  Comparison with experiment}\label{section_comparison}

In this section we apply our theory to a phosphorus donor impurity in
bulk silicon. We consider both natural samples ($f=4.67\%$ $^{29}$Si
nuclear spins) and isotopically enriched samples ($f=99.23\%$
$^{29}$Si nuclear spins). We show explicit numerical calculations of
the nuclear spin noise spectrum resulting from dipolar nuclear-nuclear
couplings, predict the Hahn echo envelope and compare our results
with the experimental data of Tyryshkin {\it et al} \cite{tyryshkin06}
and Abe {\it et al} \cite{abe04}.

\subsection{Effective mass model for a phosphorus impurity in silicon}

Here the donor impurity is described within effective mass theory by a
Kohn-Luttinger wave function \cite{kohn57},
\begin{eqnarray}
\Psi({\mathbf r}) &=& \frac{1}{\sqrt{6}}\sum_{j=1}^{6}F_{j}({\mathbf
r})u_{j}({\mathbf r})e^{i {\mathbf k_{j}\cdot r}},
\label{kohnluttinger}
\\ {\mathbf k_{j}}&=& 0.85 \frac{2\pi}{a_{\rm{Si}}}\hat{k}_{j} , \;
\hat{k}_{j} \in \left\{
\hat{x},-\hat{x},\hat{y},-\hat{y},\hat{z},-\hat{z}\right\},\\
F_{1,2}({\mathbf r}) &=& \frac{\exp{\left[
-\sqrt{\frac{x^{2}}{(nb)^{2}}+\frac{y^{2}+z^{2}}{(na)^{2}}}\right]}}
{\sqrt{\pi(na)^{2}(nb)}}\label{f12},
\end{eqnarray}
with envelope functions $F_{j}$ describing the effective mass
anisotropies. Here $n=(0.029eV/E_{i})^{1/2}$ with $E_{i}$ being the
ionization energy of the impurity ($E_{i}=0.044$ eV for the phosphorus
impurity, hence $n=0.81$ in our case), $a_{\rm{Si}}=5.43$ \AA\ the lattice
parameter for Si, $a=25.09$ \AA\ and $b=14.43$ \AA\ characteristic
lengths for Si hydrogenic impurities \cite{feher59a}. Moreover, we will
use experimentally measured values for the charge density on each Si
lattice site  $|u({\mathbf R_{i}})|^{2}=\eta\approx 186$ \cite{kohn57}.
Hence the isotropic hyperfine interaction is given by
\begin{eqnarray}
A^{\rm{iso}}_{i}&=&\frac{16\pi}{9}\gamma_{e0} \gamma_{n}\eta\left[
F_{1}({\mathbf R_{i}})\cos{(k_{0}X_{i})} \right.\nonumber\\
&&\left.+F_{3}({\mathbf R_{i}})\cos{(k_{0}Y_{i})} +F_{5}({\mathbf
R_{i}})\cos{(k_{0}Z_{i})}\right]^{2},\label{ansip}
\end{eqnarray}
with the Si conduction band minimum at $k_{0}=(0.85)2\pi/a_{\rm{Si}}$,
gyromagnetic ratios for $^{29}$Si nuclear spins $\gamma_{n}=5.31\times
10^{3}$ (sG)$^{-1}$, and the free electron $\gamma_{e0}= 1.76\times
10^{7}$ (s G)$^{-1}$.  It is instructive to check the experimental
validity of Eq. (\ref{ansip}) by calculating the inhomogeneous
line-width $\sim 1/(\gamma_{e0} T_{2}^{*})$. A simple statistical
theory [Eq.~(\ref{inhomobroad})] leads to
\begin{equation}
\langle \left( \omega/\gamma_{e0}-B\right)^{2}\rangle=
\frac{f}{(2\gamma_{e0})^{2}}\sum_{{\mathbf R_{i}}\neq {\mathbf
0}}\left(A^{\rm{iso}}_{i}\right)^2.
\label{T2star}
\end{equation}
For natural silicon (nuclear spin fraction $f=0.0467$) our calculated
root mean square line-width is equal to $0.89$ G. On the other hand, a
simple spin resonance scan leads to $2.5$ G$/2\sqrt{2\ln{2}}=1.06$
G \cite{feher59a}. Therefore the simple model employed here is able to
explain 84\% of the experimental hyperfine line-width. This is the
level of agreement that we should expect when comparing our theory for
echo decay with experiment. 

\subsection{Explicit calculations of the nuclear spin noise spectrum
  and electron spin echo decay of a phosphorus impurity in silicon}

The nuclear spin noise spectrum is calculated from
Eq.~(\ref{sijtotal}) by excluding the $\delta(\omega)$ contribution.  
For each pair $i,j$ we calculate the 
transition frequency Eq.~(\ref{epminus}) and broadening
Eq.~(\ref{sigpm}) using the derived microscopic values of the
hyperfine interaction [Eq.~(\ref{ansip})] and the dipolar interaction
\begin{equation}
b_{ij}=-\frac{1}{4}\gamma_n^2\hbar \frac{1-3\cos^{2}{\theta_{ij}}}{R_{ij}^3}.
\label{bij22}
\end{equation}
For silicon the sites $i,j$ lie in a diamond lattice with parameter
$a_{{\rm si}}=5.43$~\AA.  We wrote a computer program that sums over
lattice sites ${\bm R}_i$ within $r_0$ of the center of the donor.
Each site ${\bm R}_i$ is then summed with all sites ${\bm R}_j$ within
$r'_0$ of ${\bm R}_i$ (excluding double counting). After numerical
tests we concluded that the values $r_0=200$~\AA~and
$r'_0=10$~\AA~were high enough to guarantee convergence (increasing
$r_0$ and $r'_0$ changes the calculations by a negligible amount).
Our explicit numerical calculations for the echo decay without
broadening [Eq.~(\ref{expij})] reproduced the equivalent calculation
of Witzel {\it et al.} \cite{witzel05} with no visible deviation.  For
$k_BT\gg \gamma_n B$ we may assume that the nuclear spins are
completely unpolarized (the experimental data was taken at $T=4$~K and
$B=0.3$~T \cite{tyryshkin06}). We account for the isotopic fraction
$f$ (ratio of sites containing nuclear spin $1/2$) using a simple
averaging method. For example, the pair populations are set as
$p_+=p_-=f^2/4$, and the broadening $\sigma^{2}_{+-}\propto f$ [note
$\sum_k b^{2}_{ij}$ in Eq.~(\ref{sigpm})].

\begin{figure}[t]
\centerline{\includegraphics*[width=.7\textwidth]{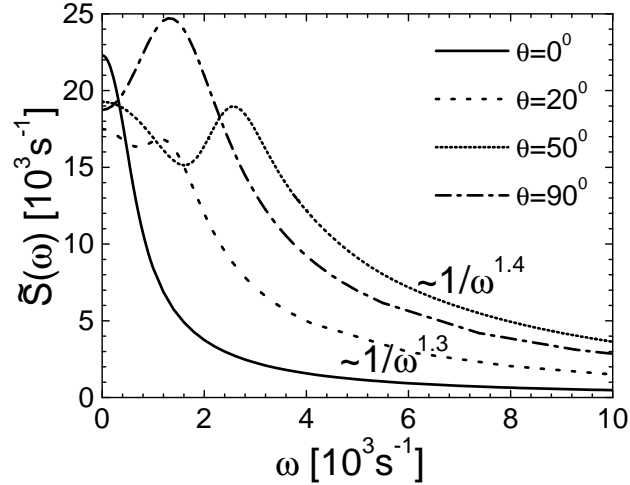}}
\caption[]{Nuclear spin noise affecting the electron spin phase. We
  show the noise spectrum for several magnetic field orientation
  angles $\theta$ with
  respect to the (001) direction. As $\theta$ is increased from zero,
  a broad peak develops at a frequency close to the dipolar splitting
  between nearest neighbors in the lattice. In this case the noise
  spectrum is clearly non-monotonous, and cannot be described by a Markovian
  model. The spin echo envelope is a frequency integral of the
  noise spectrum weighted by a filter function.}
\label{fig_noise} 
\end{figure}

Fig.~\ref{fig_noise} shows the nuclear spin noise spectrum for natural
Si at four different magnetic field orientation angles $\theta$ with
respect to the crystal direction (001). Here $\theta=0^{\circ}$
corresponds to $\bm{B}\parallel (001)$, while $\theta=90^{\circ}$
corresponds to $\bm{B}\parallel (110)$. For $\theta$ away from zero
the noise spectrum is charaterized by a broad peak at which the
flip-flop transition frequencies $E_{+-}$ accumulate. The fact that
the spectrum is non-monotonic implies important \emph{non-Markovian}
behavior for electron spin dynamics (recall that a Markovian noise
spectrum is defined as a sum of Lorentzians, hence it is always
monotonic).  Interestingly, for $\theta$ close to zero and at low
frequencies ($\omega<5\times 10^3$~s$^{-1}$), the spectrum appears to
be similar to a Lorentzian peaked at $\omega=0$ However, one can not
fit a Lorentzian up to high frequencies because the assymptotic
behavior deviates significantly from $1/\omega^2$.

\begin{figure}[t]
\centerline{\includegraphics*[width=.7\textwidth]{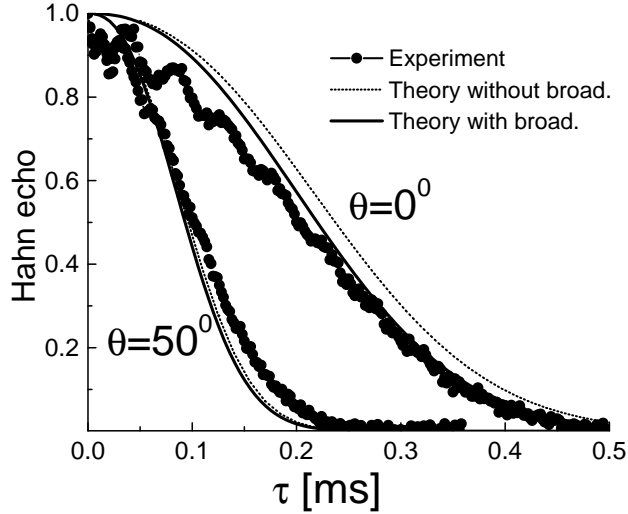}}
\caption[]{Electron spin echo decay of a phosphorus impurity in natural silicon
  (4.67\% $^{29}$Si nuclear spins) for two different magnetic field
  orientations. We show experimental data from Ref.~\cite{tyryshkin06}
  together with theoretical calculations without flip-flop broadening
  (identical to Ref.~\cite{witzel05}) and with flip-flop broadening.
  The latter is calculated by directly integrating the noise spectrum
  shown in Fig.~\ref{fig_noise} with the appropriate filter function.}
\label{fig_nat} 
\end{figure}

The Hahn echo is obtained by integrating the noise spectrum multiplied
by the filter function Eq.~(\ref{fecho}) up to a frequency cut-off
$\Lambda$ (we used $\Lambda=10^6-10^7$~s$^{-1}$, and $df\sim
1-10$~s$^{-1}$ in our numerical calculations).  The result is shown in
Fig.~\ref{fig_nat} for two different orientations. We show
calculations of the echo without broadening [Eq.~(\ref{expij}),
identical to the result shown in Ref.~\cite{witzel05}] and for the
echo with broadening, that is obtained through direct integration of
the noise spectrum shown in Fig.~\ref{fig_noise}. Note that the two
theories are in close agreement here because for low nuclear spin
density ($f=0.0467$) the broadenings are generally much smaller than
the transition frequencies $E_{+-}$, at least for the important pairs
causing spectral diffusion.  Recall that our theory does not account
for the anisotropic hyperfine interactions. Therefore our theoretical
results should be compared to the monotonic envelope enclosing the
experimental data points.\footnote{We thank Dr. A.M. Tyryshkin for
  pointing this out to us.} The echo modulations due to the
anisotropic hyperfine interaction is clearly visible at short times in
the experimental data shown in Fig.~\ref{fig_nat}. These oscillations
produce a loss of contrast of about 10\% at the short time regime.
Apart from this effect, the agreement between theory and experiment is
quite good.

\begin{figure}[t]
\centerline{\includegraphics*[width=.7\textwidth]{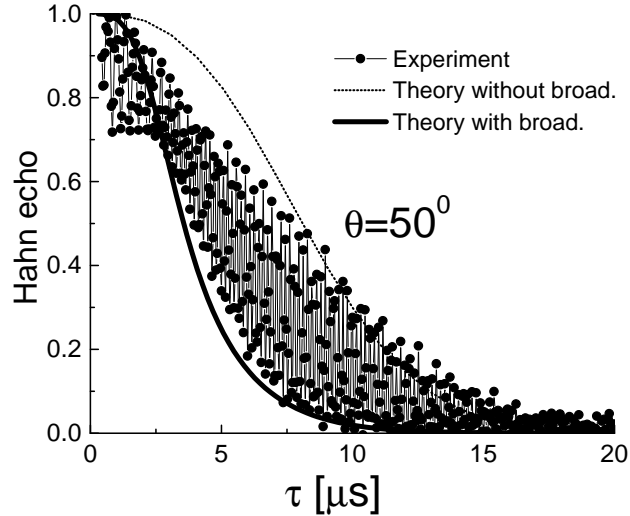}}
\caption[]{Electron spin echo decay of a phosphorus impurity in
  isotopically enriched silicon (99.23\% of $^{29}$Si).  Experimental
  data from Ref.~\cite{abe04}.}
\label{fig_si29} 
\end{figure}

Fig.~\ref{fig_si29} shows echo decay results for isotopically enriched
samples ($f=99.23\%$). The experimental data is from Abe {\it et al.}
\cite{abe04}. Note that here the echo modulations are very evident,
the loss of contrast reaches $\sim 100\%$. The monotonic envelope on
top of the experimental data is in reasonable agreement with the
theory without broadening. However, the theory with broadening decays
significantly faster. The difference between both theories increases
for increasing $f$.  This suggests that the mean-field theory proposed
in section \ref{section_broadening} overestimates the broadening. The
author expects that a more sophisticated many-body calculation may
account for this discrepancy.

\begin{figure}[t]
\centerline{\includegraphics*[width=.7\textwidth]{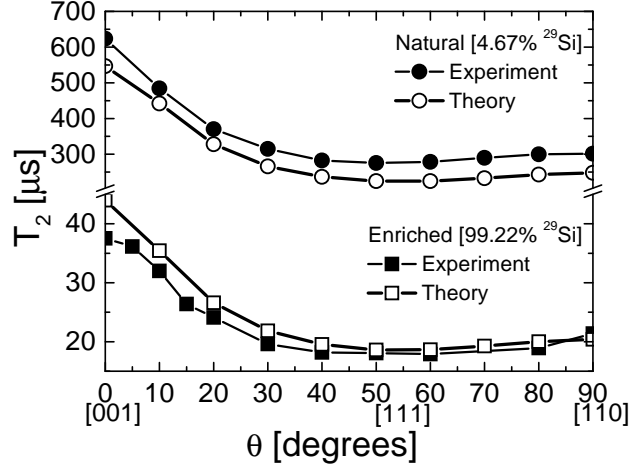}}
\caption[]{Orientation dependence of the 1/e echo decay time $T_2$. 
  $\theta$ is the angle between the applied magnetic field and the
  crystalographic (001) direction, so that $\theta=90^{\circ}$ is
  along the (110) direction.  We show experimental data for natural Si
  \cite{tyryshkin06} and for isotopically enriched Si \cite{abe04}.
  The theoretical calculations shown are without broadening. For
  natural Si, the experimental data was corrected for a $\theta$
  independent instantaneous diffusion decay, see
  Ref.~\cite{tyryshkin06}. Because of this the theoretical $T_2$'s
  are lower than the experimental $T_2$'s reported in
  Ref.~\cite{tyryshkin06}, in contrast to what is observed in
  Fig.~\ref{fig_nat}.} 
\label{fig_orientation} 
\end{figure}

Fig.~\ref{fig_orientation} shows the dependence of the $1/e$ echo
decay time ($T_2$) with the magnetic field angle. The shortest value
of $T_2$ is obtained when $\bm{B}$ is along the (111) direction
($\theta=54.74^{\circ}$). In this case none of the nearest neighbor
pairs have zero dipolar couplings. Only pairs $i,j$ with $\bm{R}_{ij}$
parallel to the (100), (010), and (001) directions have their dipolar
interaction turned off by the magic angle
[$\theta_{ij}=54.74^{\circ}$ implies $\cos{\theta_{ij}}=1/\sqrt{3}$
and $b_{ij}=0$, see Eq.~(\ref{bij22})]. On the other hand for
$\bm{B}\parallel (001)$, $T_2$ is longer by a factor of three. This
occurs because the nearest neighbor pairs, that usually give the
strongest contribution to echo decay, are forming a magic angle with
respect to $\bm{B}\parallel (001)$.\footnote{The nearest neighbors for
  each site $i$ are located at $\bm{R}_{ij}=\frac{1}{4}a_{{\rm
      Si}}(1,1,1)$, $\frac{1}{4}a_{{\rm Si}}(-1,-1,1)$,
  $\frac{1}{4}a_{{\rm Si}}(-1,1,-1)$, and $\frac{1}{4}a_{{\rm
      Si}}(1,-1,-1)$.}

We now discuss the time dependence of the echo envelope. The echo
decay without broadening fits well to the expression
\begin{equation}
\langle \sigma_+(2\tau)\rangle = 
\textrm{e}^{-\left(\frac{2\tau}{T_2}\right)^{2.3}}
\label{tdepend}
\end{equation}
for a wide range of $2\tau$ centered around $T_2$ and for all values
of $f$ (for a log-log plot, see Fig. 9 of Ref.~\cite{witzel06}).
Tyryshkin {\it et al.}  \cite{tyryshkin06} studied the time dependence
of the natural silicon experimental data by fitting the expression
$v_E(2\tau)=\textrm{e}^{-2\tau/T'_{2}}\textrm{e}^{-(2\tau/T_2)^n}$.
Here $T'_{2}$ was interpreted as arising from a combination of
spin-flip processes and the instantaneous diffusion mechanism,
due to the finite concentration of donors. Tyryshkin {\it et al.} reported
$T'_2=1.1$~ms and exponent $n=2.4\pm 0.1$ for all sample orientation
angles between $20^{\circ}\leq\theta\leq 90^{\circ}$. For
$\theta=10^{\circ}$ they found $n=2.6\pm 0.1$, while for
$\theta=0^{\circ}$ $n=3.0\pm 0.2$. The time dependence at angles close
to the (001) direction is yet to be explained theoretically. At
natural abundance ($f=0.0467$) the theory with broadening and the
theory without broadening have similar time dependences. However, as
$f$ increases the time dependence of the broadened theory deviates
significantly from the theory without broadening. As an example, for
$f=0.9923$ and $\theta=50^{\circ}$ the broadened theory shows a
cross-over from $\textrm{e}^{-\tau^{3.3}}$ at short $\tau< 3$~$\mu$s
to $\textrm{e}^{-\tau^{1.7}}$ for $\tau>3$~$\mu$s. This indicates that
adding broadening to the nuclear spin excitations leads to observable
effects in the time dependence of electron spin coherence.
Unfortunately, the echo modulations are too strong in isotopically
enriched samples (Fig.~\ref{fig_si29}).  This makes the precise
experimental determination of the time dependence of the echo envelope
quite difficult.

Eq.~(\ref{tdepend}) allows us to extract scaling of the 1/e decay time
$T_2$ with the nuclear spin fraction $f$. Note that in the theory
without broadening $f$ appears as a pre-factor in the exponent due to
$p_+ + p_-=f^2/4$. Therefore we have simply
\begin{equation}
T_{2}\propto f^{-2/2.3}=f^{-0.87}.
\label{t2f}
\end{equation}
Abe {\it et al.} \cite{abe06} measured $T_2$ for seven 
isotopically engineered 
samples with $f$ ranging from 0.2\%--100\%. Their study shows that $T_2$
must scale between $f^{-0.86}$ and $f^{-0.89}$ in good agreement with
Eq.~(\ref{t2f}).

\begin{figure}[t]
\centerline{\includegraphics*[width=.7\textwidth]{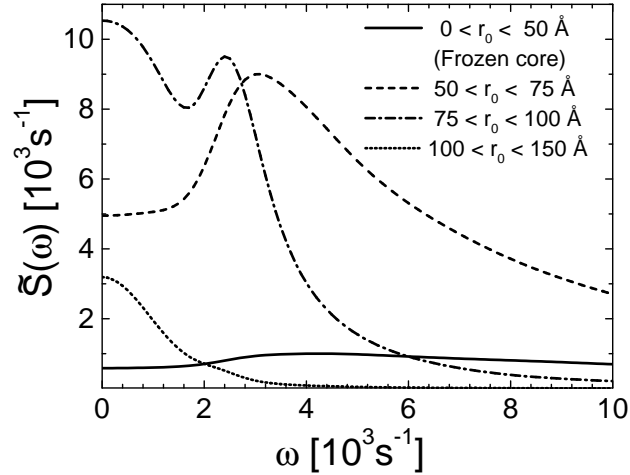}}
\caption[]{Contribution of 
  nuclear spins located at concentric shells around the donor (natural
  Si, $\theta=50^{\circ}$). Nuclear pairs closer to the center
  ($r_0<50$~\AA) have their fluctuation amplitude suppressed by the
  strong hyperfine field difference between sites $i,j$, forming a
  frozen core. The largest contribution is due to pairs located at
  $50$~\AA~$<r_0<$~$100$~\AA.}
\label{fig_dist} 
\end{figure}

It is interesting to study the number and location of nuclear spins
contributing to the noise spectrum.  Fig.~\ref{fig_dist} shows the
contribution due to pairs inside shells concentric at the donor center
(for natural silicon and $\theta=50^{\circ}$).  The contribution for
$r_0<50$~\AA~ is quite small, but extends over a wide frequency
spectrum. These nuclear spins are said to form a ``frozen core'',
because their noise amplitude is suppressed due to the strong
difference in hyperfine fields affecting sites $i,j$. The frozen core
of a Si:P donor has about $3\times 10^4$ nuclear spins.  This frozen
core effect plays an important role in other contexts as well such as
optical spectroscopy experiments \cite{devoe81}. The nuclear spin
noise theory developed here allows a quantitative description of this
effect.\footnote{In order to understand the frozen core effect from
  our analytical expression for the noise spectrum, assume $A_i\gg
  A_j$ and $\Delta_{ij}\gg b_{ij}$ in Eq.~(\ref{sigpm}). In this case
  we have $\sigma_{+-}\sim A_i$. From Eq.~(\ref{sijtotal}) the noise
  amplitude becomes $\sim b^{2}_{ij}/A_i$, that is much smaller than
  $b_{ij}$.}  From Fig.~\ref{fig_dist} it is evident that a
significant fraction of the finite frequency noise power comes from
the large number of nuclear spins located between $50$~\AA~ and
$100$~\AA~ off the donor center (about $\sim 2\times 10^5$ nuclear
spins).  These pairs are satisfying a quasi-resonance condition
$\Delta_{ij}\sim b_{ij}$.

\section{Conclusions and outlook for the future\label{section_conclusion}}

In this chapter we described the coherent evolution of an electron
spin subject to time dependent fluctuations along its quantization
axis.  We showed that in the Gaussian approximation the electron spin
transverse magnetization can be expressed as a frequency integral over
the magnetic noise spectrum multiplied by an appropriate filter
function. The filter function depends on the particular pulse sequence
used to probe spin coherence, differing substantially at low
frequencies for free induction decay (FID) and Hahn echo. For a
Gauss-Markov model (Lorentzian noise spectrum centered at zero
frequency) we showed that the short time decay of the FID signal is
approximately $\textrm{e}^{-t^2}$, while the echo decays according to
$\textrm{e}^{-\tau^3}$.

We applied this general relationship between noise and decoherence to
the case of a localized electron spin in isotopically engineered
silicon, where the magnetic noise is mainly due to the dipolar
fluctuation of spin-1/2 lattice nuclei. The nuclear spin noise
spectrum was calculated from a pair-flip-flop model, resulting in a
linear combination of sharp transitions (delta functions). The echo
decay due to these sharp transitions is identical to the one derived
by the lowest order cluster expansion \cite{witzel05}. Next, we showed
how to obtain a smooth noise spectrum by adding broadening to these
transitions using a mean-field approach.  The resulting noise spectrum
was found to be strongly non-monotonic, hence qualitatively different
from the usual Lorentzian spectrum of a Gauss-Markov model. This
structured noise spectrum is able to explain the non-Markovian
dynamics ($\sim \textrm{e}^{-\tau^{2.3}}$) observed in electron spin
echo experiments for phosphorus doped silicon. We compared the
theories with and without broadening to two sets of experimental
data, for natural and isotopically enriched silicon. The agreement was
quite good for natural silicon, but not as good for $^{29}$Si enriched
samples.

It is interesting to compare our results to the set of non-Gaussian
phenomenological theories proposed a long time ago by Klauder and
Anderson \cite{klauder62}.  These authors classified spectral
diffusion behavior in two groups, depending on the nature of the
interactions causing magnetic noise. In ``$T_1$ samples'' the magnetic
noise is caused by non-resonant spins fluctuating individually (e.g.,
due to phonon emission).  On the other hand the magnetic noise at
``$T_2$ samples'' is caused by the mutual interaction of the
non-resonant spins (For example, a nuclear spin bath weakly coupled to
the lattice is a ``$T_2$ sample'' because the longitudinal nuclear
spin relaxation time $T^{n}_{1}$ is much longer than the transverse
relaxation time $T^{n}_{2}$).  Klauder and Anderson showed that echo
decay behavior in a variety of $T_1$ samples could be described by a
Markovian theory by making assumptions about the general shape of the
distribution of fluctuations at any given time.  While a Gauss-Markov
model leads to echo decay of the form $\sim \textrm{e}^{-\tau^3}$, a
Lorentz-Markov model leads to $\textrm{e}^{-\tau^2}$ behavior, and
intermediate non-Gaussian distributions result in
$\textrm{e}^{-\tau^n}$ with $n$ between two and three. Later,
Zhidomirov and Salikhov \cite{zhidomirov69} showed that similar
behavior can be obtained in $T_1$ samples composed of a dilute
distribution of magnetic impurities fluctuating according to a random
telegraph noise model (Markovian with a non-Gaussian distribution).

Nevertheless the problem of echo decay behavior in ``$T_2$ samples''
remained open. It was found empirically by many authors (see
\cite{chiba72} and references therein) that echo decay behavior in
``$T_2$ samples'' is usually well fitted to the expression $\sim
\textrm{e}^{-\tau^2}$, and the Lorentz-Markov model of Klauder and
Anderson was often invoked as a phenomenological explanation.  Here we
show that this behavior can be derived microscopically from a Gaussian
model that takes into account the non-Markovian evolution of the
coupled nuclear spin bath. The resulting behavior found by us
[$\textrm{e}^{-\tau^{2.3}}$] is not due to a short time approximation
[the short time behavior for each pair is actually given by
$\textrm{e}^{-\tau^4}$, see Eq.~(\ref{expij})].  In order to explain
these experiments we must consider the collective finite time
evolution of a large number of nuclear spins [note that the
characteristic frequency of fluctuation for a pair flip-flop
$E_{+-}=2\sqrt{b_{ij}^{2}+\Delta_{ij}^2}$ gets renormalized to values
much larger than the dipolar interaction $b_{ij}$ when the nuclear
spins are subject to strong hyperfine inhomogeneities
$\Delta_{ij}=(A_i-A_j)/4$]. This theoretical explanation opens the way
to novel microscopic interpretations of a series of pulse electron
spin resonance experiments, where the electron spin may be viewed as a
spectrometer of low frequency magnetic noise due to a large number of
nuclear spins or other magnetic moments. For example, a recent
experiment \cite{schenkel06} revealed that magnetic noise from the
surface must play a role on spin echo decay of antimony impurities
implanted in isotopically purified silicon (with a very low density of
$^{29}$Si nuclear spins in the bulk).


There are many open questions that deserve further investigation.
First, what is the contribution of higher order nuclear spin
transitions to the noise spectrum? This question may be answered by
going beyond the simple pair flip-flop model assumed here, in a
similar fashion as the cluster expansion developed in \cite{witzel05},
or using an alternative linked cluster expansion for the spin Green's
function \cite{saikin06}.  Another interesting open question is the design
of optimal sequences for suppressing the effects of nuclear spin noise
in electron spin evolution, as was done for the random telegraph noise
model in Ref.~\cite{mottonen06}. This is particularly important in the
context of spin-based quantum computation. The efficiency of a
Carr-Purcell sequence in suppressing the electron spin coherence decay
due to a nuclear spin bath was considered both in the framework of a
semiclassical model (see Ref.~\cite{desousa05}, where the role of
nuclear spins greater than 1/2 was also considered) and using a
cluster expansion approach \cite{witzel06b}. Recently, it was shown
that the electron mediated inter-nuclear coupling
[Eq.~(\ref{veffapprox})] may be exploited in order to recover electron
spin coherence lost for the nuclear spin bath \cite{yao06}.  We will
certainly see many other interesting developments in the near future.

\subsubsection{Acknowledgements}

This research project began in 2001 in collaboration with Sankar Das
Sarma, my Ph.D. advisor. I am grateful for and appreciate his guidance
and support. The collaboration continued with Wayne Witzel, another
Ph.D. student in S. Das Sarma's group.  This part of the work was
supported by ARDA, US-ARO, US-ONR, and NSA-LPS.  My postdoctoral work
on this problem was in collaboration with K.  Birgitta Whaley and Neil
Shenvi.  Neil also dedicated part of his Ph.D.  thesis on problems
related to electron spin dynamics due to nuclear spins. This part of
the work was supported by the DARPA SPINS program and the US-ONR.  My
long, energetic and prolific collaboration with Neil and Wayne cannot
be overstated.  I acknowledge great discussions with J.  Fabian, X.
Hu, A.  Khaetskii, D.  Loss, L.J. Sham, L.M.K.  Vandersypen, and I.
\v{Z}ut\'{i}c.  The role of experimentalists in the development of
this theory was also very important. I wish to thank Stephen Lyon,
Alexei Tyryshkin, Thomas Schenkel, Kohei Itoh, and Eisuke Abe for
discussions and for providing their experimental data before it was
published.  Finally, I wish to acknowledge an enlightening
conversation with Professor Erwin Hahn, who kindly shared his
astonishing intuition on the problems that he pioneered a long time
ago.

\printindex
\end{document}